\documentclass[preprint,prd,aps,floatfix,preprintnumbers,superscriptaddress,bibnotes,nofootinbib]{revtex4-1}

\usepackage{amsmath}
\usepackage{latexsym}
\usepackage[psamsfonts]{amssymb}
\usepackage{graphicx}
\usepackage{longtable}
\usepackage{ulem}
\usepackage{longtable}
\usepackage{epstopdf}
\usepackage{bm} 
\usepackage{color}
\usepackage{hyperref}
\usepackage{multirow}

\newcommand{\be}{\begin{equation}}
\newcommand{\ee}{\end{equation}}
\newcommand{\ben}{\begin{eqnarray}}
\newcommand{\een}{\end{eqnarray}}
\newcommand{\bi}{\begin{itemize}}
\newcommand{\ei}{\end{itemize}}

\begin{document}

\preprint{UTHEP-757, UTCCS-P-138, HUPD-2106}


\title{Calculation of derivative of nucleon form factors
in $N_f = 2+1$ lattice QCD at $M_\pi = 138$ MeV on a (5.5 fm)$^3$ volume}

%
\author{Ken-Ichi~Ishikawa\:}
\affiliation{Core of Research for the Energetic Universe, Graduate School of Advanced Science and Engineering, Hiroshima University, Higashi-Hiroshima, 739-8526, Japan}
\author{Yoshinobu~Kuramashi\:}
\affiliation{Center for Computational Sciences, University of Tsukuba, Tsukuba, Ibaraki 305-8577, Japan}
\author{Shoichi~Sasaki\:}
\affiliation{Department of Physics, Tohoku University, Sendai 980-8578, Japan}
\author{Eigo~Shintani\:}
\affiliation{Center for Computational Sciences, University of Tsukuba, Tsukuba, Ibaraki 305-8577, Japan}
\author{Takeshi~Yamazaki\:}
\affiliation{Faculty of Pure and Applied Sciences, University of Tsukuba, Tsukuba, Ibaraki, 305-8571, Japan}
\affiliation{Center for Computational Sciences, University of Tsukuba, Tsukuba, Ibaraki 305-8577, Japan}

\collaboration{PACS Collaboration}

\date{\today}
\begin{abstract}
We present a direct calculation for the first derivative
of the isovector nucleon form factors with respect to the momentum transfer $q^2$
using the lower moments of the nucleon 3-point function in the coordinate space.
Our numerical simulations are performed using the $N_f = 2 + 1$ 
nonperturbatively $O(a)$-improved Wilson quark action and Iwasaki gauge action near 
the physical point, corresponding to the pion mass $M_\pi =138$ MeV, 
on a (5.5 fm)$^4$ lattice at a single lattice spacing of $a = 0.085$ fm.
In the momentum derivative approach,
we can directly evaluate the mean square radii for
the electric, magnetic, and axial-vector form factors, and also
the magnetic moment without the $q^2$ extrapolation to the zero momentum point.
These results are compared with the ones determined by the standard method, 
where the $q^2$ extrapolations of the corresponding form factors are
carried out by fitting models.
We find that the new results from the momentum derivative method 
are obtained with a larger statistical error than the standard method, 
but with a smaller systematic error associated with the data analysis.
Within the total error range of the statistical and systematic errors combined,
the two results are in good agreement.
On the other hand, two variations of the momentum derivative 
of the induced pseudoscalar form factor 
at the zero momentum point show some discrepancy. 
It seems to be caused by a finite volume effect, since 
a similar trend is not observed on a large volume, but seen on a small volume
in our pilot calculations at a heavier pion mass of $M_{\pi} = 510$ MeV.
Furthermore, we discuss an equivalence between the momentum derivative method 
and the similar approach with the point splitting vector current.
\end{abstract}


\maketitle

 
\section{Introduction}
\label{sec:introduction}

A discrepancy of experimental measurements of the proton
charge radius, called the proton radius puzzle, has not been solved yet since 
the muonic hydrogen measurement was reported in 2010~\cite{Pohl:2010zza}.
The values of the charge radius measured from both elastic 
electron-proton scattering and
hydrogen spectroscopy agree with each other~\cite{Mohr:2015ccw},
while they differ from the one measured from the muonic hydrogen.
Several experiments are carried out and also proposed 
to understand this discrepancy, see Ref.~\cite{Karr:2020wgh} 
for a review of this puzzle.
Lattice QCD calculation, which is a unique computational experiment to investigate
the complicated strong interaction dynamics, based on the first principles of QCD, 
can tackle the problem as an alternative to actual experiments.

In lattice QCD calculation, the mean square (MS) charge radius 
can be determined from the slope of 
the electric nucleon form factor at $q^2=0$.
Most calculations including 
our previous works~\cite{Ishikawa:2018rew,Shintani:2018ozy}
and recent works~\cite{Alexandrou:2017ypw,Alexandrou:2017hac,Rajan:2017lxk,Capitani:2017qpc,Green:2017keo,Alexandrou:2018sjm,Bali:2019yiy,Jang:2019vkm,Jang:2019jkn,Alexandrou:2020okk,Djukanovic:2021cgp,Park:2021ypf,Alexandrou:2021wzv}
measure the form factor in $q^2 \ne 0$ with discrete lattice momenta, 
and fit the data with appropriate functional forms, 
such as dipole form, in order to determine the form-factor slope at the zero momentum point.
The choice of fit functions, however, gives rise to relatively large systematic error. 
Even in a fit using data set including at tiny $q^2$ data point,  
its systematic error still remains as large as
the statistical error of 2\% as presented in Ref.~\cite{Shintani:2018ozy}.
Such systematic uncertainty needs to be reduced to 
draw any conclusion on a maximum discrepancy of 
about 4\% observed in the three experiments.

Apart from the systematic error from the choice of fit functions
we observed that there is a discrepancy of more than 10\%
between the experimental value and our result of 
the isovector root MS charge radius obtained from lattice QCD calculation at the physical point 
($M_\pi = 135$ MeV) on a (10.9 fm)$^3$ volume~\cite{Shintani:2018ozy}.
In our lattice QCD simulation, some systematic uncertainties stemming
from the chiral extrapolation and finite volume effect are considered to be negligible.
Furthermore, excited state contamination 
in the electric form factor is well controlled and not significant in our simulation, 
because strong dependence on the time separation between the source and sink 
operators was not observed.
A possible source of this discrepancy could
be the effect of finite lattice spacing
though it seems too large for $O(a^2)$ effect in our 
non-perturbative improved Wilson quark calculation.
Our future calculations performed at the finer lattice spacing will 
reveal the presence of the systematic error due to the lattice discretization effect.
Nevertheless, we are still pursuing other reasons. 
We are interested in recent development of another approach, 
called the momentum derivative method, which can directly calculate the slope of the form factor.
In comparison to the standard approach, this method should be useful to
pin down the source of the current discrepancy between our lattice result and the experimental value. 

The momentum derivative method
was proposed in Ref.~\cite{Aglietti:1994nx},
where the slope is determined without assuming fit functions of the form factor.
This method employs the moments of the 3-point
function in the coordinate space, which can access the derivatives of 
the form factor with respect to the square of four-momentum transfer $q^2$ at vanishing $q^2$.
This method and its variation were applied to the nucleon form factor 
with the vector current
at the pion mass of $M_\pi = 0.4$ GeV~\cite{Bouchard:2016gmc},
and the physical $M_\pi$~\cite{Alexandrou:2020aja}, respectively.
Another method of the direct derivative calculation using the point splitting 
vector current was proposed in Ref.~\cite{deDivitiis:2012vs}, 
and it was applied for the electric, magnetic, and axial-vector form factors 
at the physical $M_\pi$ in Ref.~\cite{Hasan:2017wwt}.

In this study we adopt the former method to calculate physical quantities, 
including the form-factor slopes determined at $q^2=0$ 
for both the vector and axial-vector channels, 
in (2+1)-flavor lattice QCD at very close to the physical $M_\pi$ on the (5.5 fm)$^3$ spatial volume.
The physical quantities obtained from the momentum derivative method 
are compared with those from the standard analysis for the form factors.
Using similar simulation setup as described in our previous works~\cite{Ishikawa:2018rew,Shintani:2018ozy},
possible systematic errors involved in the momentum derivative method are discussed
by examining the effect of excited state contaminations with three 
source-sink separations and also by the finite volume study with two different volumes
in our pilot calculations at a heavier pion mass of $M_\pi = 0.51$ GeV.
In this paper we also elucidate the equivalence between the above-mentioned two direct methods 
through the discussion of an infinitesimal transformation on the correlation functions.
This study is regarded as a feasibility study towards more realistic
calculation with the PACS10 configurations on the (10.9 fm)$^3$ volume~\cite{Shintani:2018ozy,Ishikawa:2018jee,Ishikawa:2019qwn}.

This paper is organized as follows.
Section~\ref{sec:methods} explains definitions for
the nucleon correlation functions and their derivative calculated by 
moments of the correlation functions used in this study.
We also discuss the equality between two types of the direct methods, 
that were proposed in Ref.~\cite{Aglietti:1994nx} and Ref.~\cite{Hasan:2017wwt},
to calculate the derivative of the form factors in this section.
The simulation parameters are described in Sec.~\ref{sec:calc_param}.
The results from the derivative of nucleon correlation functions are
presented in Sec.~\ref{sec:result}.
Section~\ref{sec:summary} is devoted to summary of this study.
In two appendices, we firstly describe how the momentum derivative method is associated 
with a partially quenched approximation and secondly summarize the results obtained from the 
standard analysis of the form factors. 

All dimensionful quantities are expressed in units of 
the lattice spacing throughout this paper, 
unless otherwise explicitly specified. A bold-faced variable represents a three-dimensional vector.

\section{Calculation methods}
\label{sec:methods}

\subsection{Correlation function with momentum}

The exponentially smeared quark operator $q_S(t,{\bf x})$
with the Coulomb gauge fixing
is employed in this study to calculate the nucleon 2- and 3-point functions as
\begin{equation}
q_S(t,{\bf x}) = \sum_{\bf y} \phi(|{\bf y}-{\bf x}|) q(t,{\bf y}) ,
\label{eq:smeared_quark}
\end{equation}
where $q(t,{\bf x})$ presents a local quark operator, and the color and Dirac indexes are omitted.
A smearing function $\phi(r)$ is given in a spatial extent of $L$ as
\begin{equation}
\phi(r) = \left\{
\begin{array}{ll}
1 & (r=0)\\
A e^{-Br} & (r < L/2)\\
0 & (r \ge L/2)
\end{array}
\right. .
\end{equation}
with two parameters $A$ and $B$.
The nucleon 2-point function with the local sink operator is defined as
\begin{equation}
C_{LS}(t-t_{\rm src},{\bf x}-{\bf x}_{\rm src}) = 
\frac{1}{4}{\rm Tr}\left[{\mathcal P}_+
\langle 0| 
N_L(t,{\bf x}) \overline{N}_S(t_{\rm src},{\bf x}_{\rm src})
| 0\rangle\right] ,
\end{equation}
where ${\mathcal P}_+ = (1+\gamma_4)/2$, and the nucleon operator is given 
for the proton state by
\begin{equation}
N_L(t,{\bf x}) = \epsilon_{abc} u^T_a(t,{\bf x}) C\gamma_5 d_b(t,{\bf x})
u_c(t,{\bf x}) 
\end{equation}
with $C=\gamma_4\gamma_2$, the up and down quark operators $u,d$, and
$a,b,c$ being the color indexes.
The smeared source operator $N_S(t,{\bf x})$ is the same as 
the local one $N_L(t,{\bf x})$, but all the quark operators $u,d$
are replaced by the smeared ones defined in Eq.~(\ref{eq:smeared_quark}).
We also calculate the smeared sink 2-point function $C_{SS}(t,{\bf x})$,
where $N_L(t,{\bf x})$ is replaced by $N_S(t,{\bf x})$.
The momentum projected 2-point function is then given by
\begin{equation}
C_{XS}(t;{\bf p}) 
= \sum_{\bf r} {e^{-i{\bf p}\cdot{\bf r}}}
C_{XS}(t,{\bf r}) ,
\label{eq:2pt_mom}
\end{equation}
with $X = L, S$ and a three-dimensional momentum ${\bf p}$.
In a large $t$ region, the 2-point function behaves as a single exponential
function,
\begin{equation}
C_{XS}(t;{\bf p}) = \frac{E_N(p)+M_N}{2E_N(p)} Z_X(p) Z_S(p) e^{-t E_N(p)},
\label{eq:2pt_t-dep}
\end{equation}
where $M_N$ and $E_N(p)$ are the nucleon mass and energy
with the momentum $p \equiv |{\bf p}|$, respectively.
The overlap of the nucleon operator to the nucleon state is defined by
$\langle 0 | N_X(0,{\bf 0}) | N(p) \rangle = Z_X(p) u_N(p)$,
where $u_N(p)$ is a nucleon spinor.

We evaluate the nucleon 3-point functions as 
\begin{equation}
C^{k}_{{\mathcal O}_\alpha}
(t-t_{\rm src},{\bf x}-{\bf x}_{\rm src}) =
\frac{1}{4}\sum_{\bf y}
{\rm Tr}\left[
{\mathcal P}_k \langle 0 | 
N_S(t_{\rm sink},{\bf y})
J^{\mathcal O}_\alpha(t,{\bf x})
\overline{N}_S(t_{\rm src},{\bf x}_{\rm src})
| 0 \rangle \right] ,
\label{eq:3pt_x-dep}
\end{equation}
where ${\mathcal P}_k$ is a projection operator for 
${\mathcal P}_{t} = {\cal P}_+$ and
${\mathcal P}_{5j} = {\mathcal P}_+ \gamma_5 \gamma_j$ for $j=x,y,z$,
and $J^{\mathcal O}_\alpha$ is an isovector local current operator as
$J^{\mathcal O}_\alpha = \overline{u}{\mathcal O}_\alpha u -
\overline{d}{\mathcal O}_\alpha d$
with ${\mathcal O}_\alpha = \gamma_\alpha, \gamma_\alpha\gamma_5$ for
the vector ($V$) and axial-vector ($A$) currents, respectively.

The form factors in non-zero momentum transfers are calculated by
the momentum projected 3-point function as
\begin{equation}
C^{k}_{{\mathcal O}_\alpha}(t;{\bf p}) =
\sum_{{\bf r}} e^{-i{\bf p}\cdot{\bf r}}
C^{k}_{{\mathcal O}_\alpha}(t,{\bf r}) .
\end{equation}
We evaluate three types of 3-point function,
$C^t_{V_t}(t;{\bf p})$, $C^{5j}_{V_i}(t;{\bf p})$, and 
$C^{5j}_{A_i}(t;{\bf p})$ for $i = x,y,z$
to obtain the electric and magnetic form factors, $G_E(q^2), G_M(q^2)$,
and the axial-vector and induced pseudoscalar form factors, 
$F_A(q^2), F_P(q^2)$.

The asymptotic forms in $t \gg t_{\rm src}$ and
$t_{\rm sink} \gg t$, where $t_{\rm src}$ and $t_{\rm sink}$ are
defined in Eq.~(\ref{eq:3pt_x-dep}), for each 3-point function are
given by
\begin{eqnarray}
C^{t}_{V_t}(t;{\bf p}) &=& 
\frac{1}{Z_V}\widetilde{C}_2(t;{\bf p})e^{-M(t_{\rm sink}-t)}G_E(q^2) ,
\label{eq:C_GE}\\
C^{5j}_{V_i}(t;{\bf p}) &=& 
\frac{1}{Z_V}\widetilde{C}_2(t;{\bf p})e^{-M(t_{\rm sink}-t)}
\frac{i\varepsilon_{ijk}q_k}{E_N(p)+M}G_M(q^2) ,
\label{eq:C_GM}\\
C^{5j}_{A_i}(t;{\bf p}) &=& 
\frac{1}{Z_A}\widetilde{C}_2(t;{\bf p})e^{-M(t_{\rm sink}-t)}
\left[F_A(q^2)\delta_{ij}-\frac{q_iq_j}{E_N(p)+M_N}F_P(q^2)\right] ,
\label{eq:C_FA_FP}
\end{eqnarray}
where the squared momentum transfer 
is given by $q^2 = 2M_N(E_N(p)-M_N)$ with ${\bf q} = {\bf p}$.
The renormalization factors $Z_V$ and $Z_A$ are defined through the renormalization 
of the local vector and axial-vector currents on the lattice, respectively.
All 3-point functions share a common part of $\widetilde{C}_2(t;{\bf p})$, which is similar to 
the asymptotic form of the 2-point function in Eq.~(\ref{eq:2pt_t-dep}),
\begin{equation}
\widetilde{C}_2(t;{\bf p}) =
\frac{E_N(p)+M_N}{2E_N(p)} Z_S(0) Z_S(p) e^{-t E_N(p)} .
\label{eq:2pt_mod}
\end{equation}
However, $\widetilde{C}_2(t;{\bf p})$ is simply eliminated by considering the 
following ratio~\cite{Gockeler:2003ay,Hagler:2003jd}
\begin{equation}
   {\cal R}^{k}_{{\cal O}_\alpha}(t;{\bm p})= 
   \frac{C_{{\cal O}_\alpha}^{k}(t;{\bm p})}{C_{SS}(t_{\rm sink};{\bm 0})}
   \sqrt{
     \frac{C_{LS}(t_{\rm sink}-t;{\bm p})C_{SS}(t;{\bm 0})
     C_{LS}(t_{\rm sink};{\bm 0})}
     {C_{LS}(t_{\rm sink}-t; {\bm 0})C_{SS}(t;{\bm p})
     C_{LS}(t_{\rm sink}; {\bm p})}
},
   \label{eq:RatioQ}
\end{equation}
which is constructed from a given 3-point function defined in Eqs.~(\ref{eq:C_GE})--(\ref{eq:C_FA_FP}) 
with appropriate combination of 2-point functions (\ref{eq:2pt_t-dep}).
The ratios for each 3-point function give the following asymptotic values in the asymptotic region:
\begin{eqnarray}
{\cal R}^{t}_{V_t}(t;{\bm p}) &=& 
\frac{1}{Z_V}
\sqrt{\frac{E_N(p)+M_N}{2E_N(p)}}G_E(q^2),\label{eq:R_GE}\\
{\cal R}^{5j}_{V_i}(t;{\bm p})
&=& \frac{1}{Z_V}
\frac{i\varepsilon_{ijk}q_k}{\sqrt{2E_N(p)(E_N(p)+M_N)}}G_M(q^2),\label{eq:R_GM}\\
{\cal R}^{5j}_{A_i}(t,{\bm p})
&=& \frac{1}{Z_A}
\sqrt{\frac{E_N(p)+M_N}{2E_N(p)}}\left[F_A(q^2)\delta_{ij}-\frac{q_iq_j}{E_N(p)+M_N}F_P(q^2)\right] 
 \label{eq:R_FA_FP} ,
\end{eqnarray}
which contain the respective form factors.

The MS radius of a form factor ${\cal G}_O(q^2)$ is defined by
\begin{equation}
\langle r^2_O \rangle = 
\left. -\frac{6}{{\cal G}_O(0)}\frac{d {\cal G}_O(q^2)}{d q^2} \right|_{q^2=0} 
\end{equation}
with ${\cal G}_O = G_E, G_M, F_A$.
In the standard way to determine the MS radius, ${\cal G}_O(q^2)$ is at first fitted by
dipole, quadratic, and z-expansion forms~\cite{Boyd:1995cf,Hill:2010yb} 
given by
\begin{eqnarray}
{\cal G}_O(q^2) &=& \frac{{\cal G}_O^d(0)}{(1+c^d q^2)^2}\ \ 
{\rm (dipole)}, 
\label{eq:dipole_form}\\
&=& {\cal G}_O^q(0) + c_1^q q^2 + c_2^q q^4\ \ {\rm (quadratic)} , 
\label{eq:quad_form}\\
&=& {\cal G}_O^z(0) + c_1^z z(q^2) + c_2^z z(q^2)^2 + c_3^z z(q^2)^3 
\ \ (\text{z-expansion}) ,
\label{eq:z-exp_form}
\end{eqnarray}
where the z-expansion makes use of a conformal mapping from $q^2$ 
to a new variable $z$ defined as
\begin{equation}
z(q^2)=\frac{\sqrt{t_{\rm cut}+q^2}-\sqrt{t_{\rm cut}}}{\sqrt{t_{\rm cut}+q^2}+\sqrt{t_{\rm cut}}}
\end{equation}
with $t_{\rm cut}=4M_{\pi}^2$
for $G_E$ and $G_M$, or with $t_{\rm cut}=9M_{\pi}^2$ for $F_A$, 
where $M_{\pi}$ corresponds to the simulated pion mass.
Thanks to its rapid convergence of Taylor's series expansion in terms of $z$, 
we employ a cubic z-expansion form as a model independent fit as described
in our previous work~\cite{Shintani:2018ozy}.
Using the resulting fit parameter given in each fit, the MS radius can be determined as
\begin{eqnarray}
\langle r^2_O \rangle &=& 
\frac{12}{c^d}
=
-\frac{6}{{\cal G}_O^q(0)}c_1^q
=
-\frac{6}{{\cal G}_O^z(0)}\frac{c_1^z}{4 t_{\rm cut}} .
\end{eqnarray}

\subsection{Momentum derivatives of the 2- and 3-point functions}
\label{sec:der_corr}

As proposed in Ref.~\cite{Aglietti:1994nx}, 
the second-order momentum derivative of the 2-point function with respective to $p_i$
at the zero momentum point is calculated by
\begin{equation}
C_{LS}^{(2)}(t) = -\sum_{\bf r}r_i^2 C_{LS}(t,{\bf r}) ,
\label{eq:def_der_2pt}
\end{equation}
where the summation is calculated over $-L/2+1 \le r_i \le L/2$.
The superscript (2) in $C_{LS}^{(2)}$ denotes the second-order derivative. 
In a large $t$ region, a ratio of the derivative function at the zero momentum point to the zero momentum 
2-point function becomes
\begin{equation}
R_2(t) = \frac{C_{LS}^{(2)}(t)}{C_{LS}(t;{\bf 0})}
= A + \frac{t}{M_N} ,
\label{eq:asympt_rat_2pt}
\end{equation}
where the first term $A$ represents 
a constant contribution corresponding to the derivative of the amplitude
in Eq.~(\ref{eq:2pt_t-dep}).
It should be noted that the constant $A$ does not contain
both the overlap with the local operator, $Z_L$, and its derivative. 
This is simply because $Z_X(p)$ becomes independent of $p$ for the local operator ($X=L$).

The derivative of the 3-point function with respective to the momentum
is calculated in the same way as the one of the 2-point function in Eq.~(\ref{eq:def_der_2pt}).
We shall call the method to calculate the momentum derivatives of the 3-point functions as the 
derivative of form factor (DFF) method in the following.
For the vector current, 
we construct the first, second and third-order derivatives of the 3-point function 
with the appropriate type of
3-point functions, $C^{t}_{V_t}(t,{\bf r})$ or $C^{5j}_{V_i}(t,{\bf r})$ as
\begin{eqnarray}
C^{5j,(1)}_{V_i,(k)}(t) &=& -i\sum_{\bf r} r_k
C^{5j}_{V_i}(t,{\bf r}) ,
\label{eq:der_vi_1}\\
C^{t,(2)}_{V_t,(l)}(t) &=& -\sum_{\bf r}r^2_l
C^{t}_{V_t}(t,{\bf r}) ,
\label{eq:der_v4_2}\\
C^{5j,(3)}_{V_i,(kl)}(t) &=& i\sum_{\bf r}r_k r^2_l 
C^{5j}_{V_i}(t,{\bf r}) 
\label{eq:der_vi_3},
\end{eqnarray}
with $k \ne i \ne j$ and $l = x,y,z$.
Respective ratios associated with the first to third-order derivatives are defined as
\begin{equation}
R_{V_\alpha,(l)}^{k,(n)}(t) = 
\frac{C^{k,(n)}_{V_\alpha,(l)}(t)}{C^{t}_{V_t}(t;{\bf 0})} .
\label{eq:rat_der_v} 
\end{equation}
with the vector three-point function with the zero momentum, $C^{t}_{V_t}(t;{\bf 0})$. 
The superscript $(n)$ denotes the $n$-th order derivative.

In this study, we will later determine the magnetic moment, MS charge radius, and MS magnetic radius from the ratios associated with the first-order derivative~(\ref{eq:der_vi_1}), second-order derivative~(\ref{eq:der_v4_2}) 
and third-order derivative~(\ref{eq:der_vi_3}), respectively, 
without the $q^2$ extrapolations of the corresponding form factors toward the zero momentum transfer.

For the axial-vector current, the following three types of the second-order derivatives are considered in the DFF method,
\begin{eqnarray}
C^{5j,(2)}_{A_j, (i)}(t) &=& -\sum_{\bf r}r_i^2
C^{5j}_{A_j}(t,{\bf r}) ,
\label{eq:der_a3_2_i}\\
C^{5j,(2)}_{A_j, (j)}(t) &=& -\sum_{\bf r}r_j^2
C^{5j}_{A_j}(t,{\bf r}) ,
\label{eq:der_a3_2_j}\\
C^{5j,(2)}_{A_i, (ij)}(t) &=& -\sum_{\bf r}r_ir_j
C^{5j}_{A_i}(t,{\bf r}) 
\label{eq:der_ai_2},
\end{eqnarray}
which are defined with $i \ne j$.
Just as in the vector cases, the ratios associated with three types of the
second-order derivatives are evaluated as below, 
to directly access the MS axial-vector radius and the value of $F_P(0)/g_A$ with the axial-vector coupling $g_A$,
\begin{equation}
R_{A_\alpha, (l)}^{k,(2)}(t) = 
\frac{C^{k,(2)}_{A_\alpha, (l)}(t)}{C^{5j}_{A_j}(t;{\bf 0})}
\label{eq:rat_der_a} 
\end{equation}
with $l = i,j,ij$.

In the asymptotic region of $t \gg t_{\rm src}$ and $t_{\rm sink}\gg t$, 
the ratios, which are associated with the $n$-th order derivatives defined
in Eqs.~(\ref{eq:rat_der_v}) and (\ref{eq:rat_der_a}), for $n\ge 2$, exhibit
the following asymptotic behavior
\begin{equation}
R_{{\cal O}_\alpha, (l)}^{k,(n)}(t) = C + A + \frac{t}{M_N} ,
\label{eq:asympt_rat_3pt}
\end{equation}
where the first term $C$ represents a constant contribution that contains the MS radius of the form factor.
The second and third terms can be identified with the contributions from
the second-order derivative of $-\widetilde{C}_2(t;{\bf p})/\widetilde{C}_2(t;{\bf 0})$ 
in Eq.~(\ref{eq:2pt_mod}),
whose values coincide with the ones evaluated from $R_2(t)$ defined in Eq.~(\ref{eq:asympt_rat_2pt}).
Thus, the constant $C$ can be isolated using $R_2(t)$ 
to subtract the other two contributions from Eq. (\ref{eq:asympt_rat_3pt}).

\subsection{Equivalence on two definitions of momentum derivatives}
\label{sec:def_mom_derivative}

In this subsection we intend to discuss an equivalence of the two direct derivative methods.
In the following discussion, variables $x,y,z$ represent four-dimensional coordinates.

The momentum derivative of the quark propagator at the zero momentum
is given by
\begin{equation}
\left.\frac{\partial G(x,y)e^{i{\bf p}\cdot({\bf x}-{\bf y})}}{\partial p_j}
\right|_{{\bf p}=0} = 
-i(x_j-y_j) G(x,y) ,
\label{eq:def_mom_der_xy}
\end{equation}
where $G(x,y)$ represents the quark propagator.
This definition is employed in our calculation and 
Refs.~\cite{Aglietti:1994nx,Lellouch:1994zu,Feng:2019geu,Alexandrou:2020aja}.
Another definition~\cite{deDivitiis:2012vs} of the momentum derivative
is given by means of the point splitting vector current $\widetilde{\gamma}_j$ as
\begin{equation}
\left.\frac{\partial G_p(x,y)}{\partial p_j}\right|_{{\bf p}=0} = 
-i\sum_z G(x,z)\tilde{\gamma}_j(z) G(z,y) ,
\label{eq:def_mom_der_vec}
\end{equation}
where $G_p(x,y)$ represents the quark
propagator calculated through the phase rotation of
the gauge link as $U_j(x)\rightarrow e^{ip_j} U_j(x)$ with the phase associated with the momentum $p_j$.
The definition of $\widetilde{\gamma}_j$ appearing in Eq.(\ref{eq:def_mom_der_vec}) is given by
\begin{equation}
f(x) \widetilde{\gamma}_\mu(x) g(x) = \frac{1}{2}\left[
f(x+\hat{\mu})(1+\gamma_\mu)U_\mu^\dagger(x)g(x)-
f(x)(1-\gamma_\mu)U_\mu(x)g(x+\hat{\mu})\right] .
\end{equation}
We will discuss an equivalence of the above two definitions as below.

Let us consider
an infinitesimal transformation of the quark field $\psi(x) \to (1+i\alpha(x))\psi(x)$
with an arbitrary infinitesimal parameter $\alpha(x)$ depending on $x$.
Requiring the invariance of the expectation value of the quark propagator 
under this transformation, one finds the following relation:
\begin{eqnarray}
\delta(\langle \psi(x)\overline{\psi}(y)\rangle) & = &
-i\sum_z (\partial^f_\mu \alpha(z)) 
\langle \widetilde{V}_\mu (z) \psi(x)\overline{\psi}(y)\rangle 
+ i(\alpha(x)-\alpha(y))\langle \psi(x)\overline{\psi}(y)\rangle
 =  0 ,
\end{eqnarray}
where 
$\widetilde{V}_\mu (z) = \overline{\psi}(z)\widetilde{\gamma}_\mu(z)\psi(z)$
corresponds to the conserved vector current, and $\partial_\mu^f$ represents a forward difference.
When $\alpha(x) = \varepsilon x_j$ with $\varepsilon \ne 0$, we obtain
\begin{equation}
-i\sum_z \langle \widetilde{V}_j (z)
\psi(x) \overline{\psi}(y)\rangle
= -i(x_j - y_j)\langle \psi(x)\overline{\psi}(y)\rangle .
\label{eq:equiv_mom_der}
\end{equation}
The right hand side of Eq.~(\ref{eq:equiv_mom_der}) coincides with the definition of the momentum derivative
of the propagator used in our calculation, while the left hand side of Eq.~(\ref{eq:equiv_mom_der}) 
can be identified with the momentum derivative defined by Eq.~(\ref{eq:def_mom_der_vec}).
This discussion can be easily applied to hadronic correlators,
and also extended to the higher-order derivatives by using the transformation repeatedly.
The similar transformation can be applied to the temporal direction as well.
Note that the left hand side of Eq.~(\ref{eq:equiv_mom_der}) receives contribution from the quark
disconnected diagram in general, though it is absent if the isovector current is considered,
or it can be neglected if a partially quenched approximation is applied.
The former case will be explained below with the nucleon 2-point function,
and the details of the latter case are described in Appendix~\ref{app:pq_approx}.

Here let us consider the proton 2-point function with the local operators having {\it exact isospin symmetry}.
Performing the transformation of the $u$ quark fields in the
2-point function, one obtains the following relation
\begin{equation}
\langle 0 | N_L(t,{\bf x}) 
\sum_z \widetilde{V}^u_\nu(z) \overline{N}_L(0,{\bf 0}) 
| 0 \rangle
= 2 x_\nu \langle 0 | N_L(t,{\bf x}) \overline{N}_L(0,{\bf 0}) | 0 \rangle ,
\end{equation}
where $\widetilde{V}^u_\nu$ is the conserved vector current for the $u$ quark.
The factor of 2 comes from the fact that the proton operator
has two $u$ quark fields.
A similar relation is obtained through the same transformation of 
the $d$ quark fields as
\begin{equation}
\langle 0 | N_L(t,{\bf x}) 
\sum_z \widetilde{V}^d_\nu(z) \overline{N}_L(0,{\bf 0}) 
| 0 \rangle
= x_\nu \langle 0 | N_L(t,{\bf x}) \overline{N}_L(0,{\bf 0}) | 0 \rangle ,
\end{equation}
where $\widetilde{V}^d_\nu$ is the conserved vector current for the $d$ quark.
As for the isovector current 
$\widetilde{V}^v_\nu(z) = \overline{u}(z)\widetilde{\gamma}_\nu(z) u(z) -
\overline{d}(z)\widetilde{\gamma}_\nu(z) d(z)$, the corresponding relation is given by a difference of 
the above two relations as,
\begin{equation}
\langle 0 | N_L(t,{\bf x}) 
\sum_z \widetilde{V}^v_\nu(z) \overline{N}_L(0,{\bf 0}) 
| 0 \rangle
= x_\nu \langle 0 | N_L(t,{\bf x}) \overline{N}_L(0,{\bf 0}) | 0 \rangle, 
\label{eq:equiv_isovector}
\end{equation}
where there is no contribution from quark disconnected diagrams in the left hand side, since
they can be canceled due to exact isospin symmetry.

In our pilot calculations at a heavier pion mass of $M_\pi = 0.51$ GeV,
we verify the left and right equality of the above equation~(\ref{eq:equiv_isovector}), {\it i.e.,}
the equivalence between the two definitions of the momentum derivative.
We numerically confirm that they are reasonably consistent with
each other through verification of the following two equations as,
\begin{eqnarray}
{\rm Tr}\left[{\mathcal P}_+
\sum_{\bf x}\langle 0 | N_S(t,{\bf x}) 
\sum_z \widetilde{V}^v_t(z) \overline{N}_S(0,{\bf 0}) 
| 0 \rangle \right]
&=& t {\rm Tr}\left[{\mathcal P}_+
\sum_{\bf x}\langle 0 | N_S(t,{\bf x}) \overline{N}_S(0,{\bf 0}) | 0 \rangle
\right] , \nonumber \\
\label{eq:der_equiv1}\\
{\rm Tr}\left[{\mathcal P}_+ \sum_{\bf x} x_i \langle 0 | N_L(t,{\bf x}) 
\sum_z \widetilde{V}^v_i(z)\overline{N}_L(0,{\bf 0}) 
| 0 \rangle \right]
&=&
{\rm Tr}\left[{\mathcal P}_+ \sum_{\bf x} x_i^2 \langle 0 | 
N_L(t,{\bf x}) \overline{N}_L(0,{\bf 0}) | 0 \rangle \right] .\nonumber\\
\label{eq:der_equiv2}
\end{eqnarray}
The numerical verification of Eq.~(\ref{eq:der_equiv1}) 
is essentially the same as that of the renormalization factor of $Z_{\widetilde{V}} = 1$.
It should be noted that the equality described in Eq.~(\ref{eq:der_equiv2})
is valid only if both the 2-point and 3-point functions are constructed
by the {\it local} nucleon operators $N_L$. 
If the smeared nucleon operators $N_S$ are used, 
the corresponding relation becomes more complicated due to explicit spatial 
dependence of the smearing function, 
{\it i.e.,} $\phi(r)$ appearing in Eq.~(\ref{eq:smeared_quark}).
On the other hand, the equality described in Eq.~(\ref{eq:der_equiv1})
can be applicable even for the smeared nucleon operators $N_S$, since
the coordinate in the direction of the derivative is not spatial but temporal. 
The smearing function $\phi(r)$ is clearly independent of the temporal coordinate.

Finally, recall that
the smeared nucleon source operator is adopted in our whole calculation. 
Thus, the quantity, which we actually calculated as defined in Eq.~(\ref{eq:def_der_2pt}), may
exhibit a slight difference from the right hand side of Eq.~(\ref{eq:der_equiv2}) in the asymptotic region.
This difference is absorbed only into the difference of the constant $A$, which
is attributed to the difference in the overlaps to the nucleon, $Z_X(p)$, 
defined in Eq.~(\ref{eq:2pt_t-dep}), and also their derivatives, between $X=L$ and $S$. 
Therefore, it should not matter for extraction of the physical quantities using the DFF method.

Hereafter, we adopt the definition of Eq.~(\ref{eq:def_mom_der_xy}) for the momentum derivative 
in our whole calculation, since the calculation cost of the right hand side in Eqs.~(\ref{eq:der_equiv1}) 
and (\ref{eq:der_equiv2}) is much smaller than that of the left hand side.

\section{Calculation parameters}
\label{sec:calc_param}

The configuration used in this study
was generated with the Iwasaki gauge action~\cite{Iwasaki:2011jk}
and the six stout-smeared Clover quark action near the physical point
for intended purpose where the finite-size study was done for 
light hadron spectroscopy in $N_f=2+1$ lattice QCD 
at the physical point~\cite{Ishikawa:2018jee,Ishikawa:2019qwn}.
The lattice size is $L^3\times T = 64^3\times 64$
corresponding to (5.5 fm)$^4$ with 
a lattice cutoff, $a^{-1} = 2.3162(44)$ GeV~\cite{Ishikawa:2019qwn}.
Details of the parameters for the gauge configuration generation are summarized
in Ref.~\cite{Ishikawa:2018jee}.

\begin{figure}[!ht]
 \centering
\includegraphics*[scale=0.35]{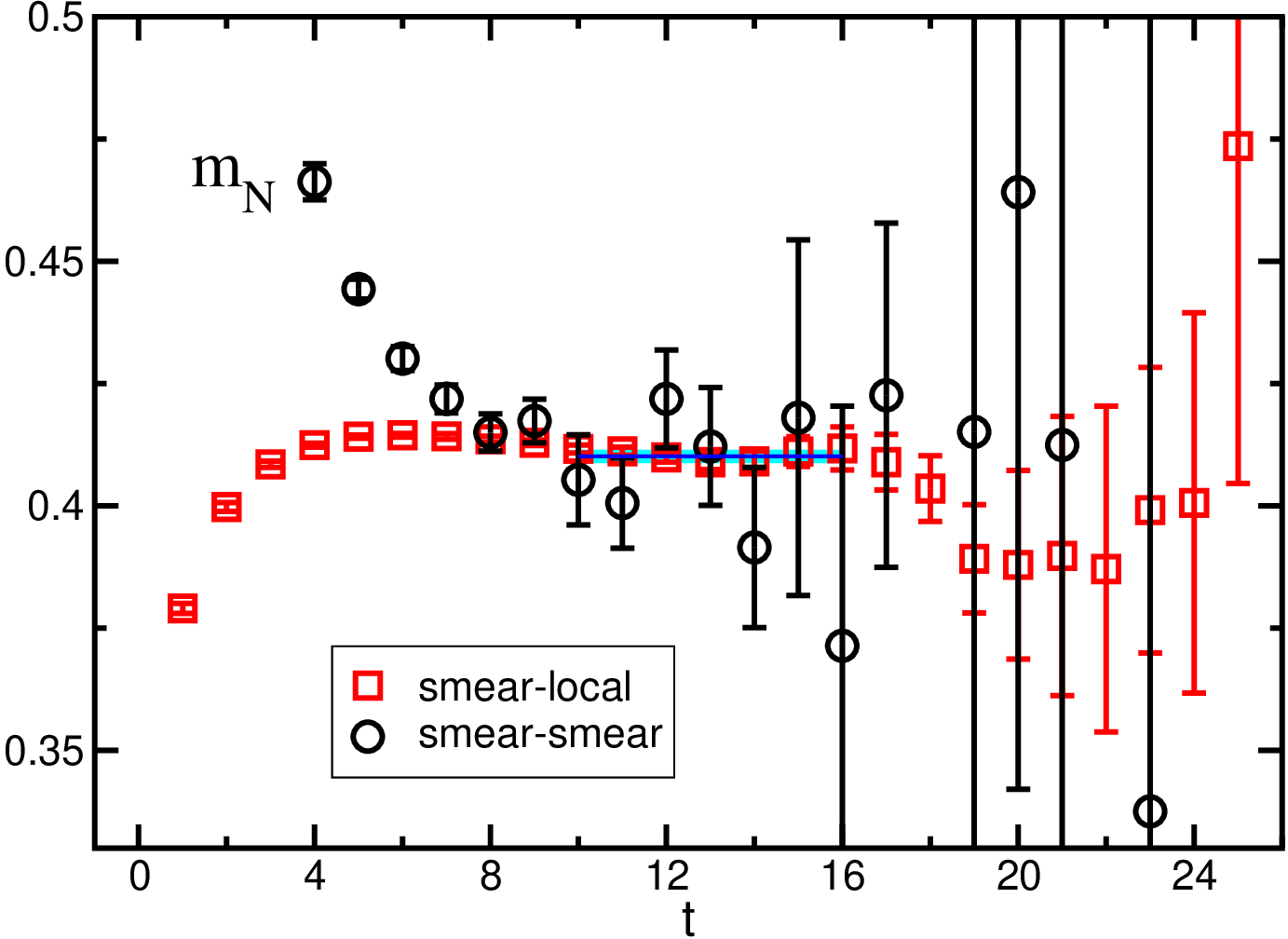}
 \caption{Effective mass for the nucleon from the smear-local (squared symbols)
 and smear-smear (circle symbols) cases of the nucleon 2-point functions, $C_{LS}(t;{\bf 0})$ 
 and $C_{SS}(t;{\bf 0})$.
 Horizontal solid line represents a value of the nucleon mass obtained from $C_{LS}(t;{\bf 0})$ 
 (smear-local) by a single exponential fit. Shaded band indicates its statistical error 
 and the fitting range.
  \label{fig:eff_N}
 }
\end{figure}

For the measurements for the nucleon correlation functions,
the same quark action as in the gauge configuration generation
is employed with the hopping parameter $\kappa = 0.126117$ for the light quarks
and the improved coefficient $c_{\rm SW} = 1.11$~\cite{Taniguchi:2012kk}.
The quark propagator is calculated using the exponential smeared source
in Eq.~(\ref{eq:smeared_quark}) with the Coulomb gauge fixing.
The smearing parameters for the quark propagator are 
chosen as $(A,B) = (1.2,0.14)$ to obtain early plateau of the effective mass 
of $C_{LS}(t;{\bf 0})$ as shown in Fig.~\ref{fig:eff_N}.
The periodic boundary condition in all the temporal and spatial directions is
adopted in the quark propagator calculation.
The sequential source method is used to calculate 
the nucleon 3-point functions with 
$t_{\rm sep} = t_{\rm sink}-t_{\rm src} = 12, 14,$ and 16
corresponding to 1.02, 1.19, and 1.36 fm, respectively.
Our main result is obtained with $t_{\rm sep}=14$,
and the results of $t_{\rm sep}=12$ and 16 are used for comparison.
These values of $t_{\rm sep}$ are the same as the ones used
in our previous calculation~\cite{Shintani:2018ozy},
where significant excited state contributions were {\it not observed} in
three particular form factors of $G_E$, $G_M$, and $F_A$.

The nucleon 2- and 3-point functions are measured with
the 100 configurations separated by the 20 molecular dynamics trajectories.
Their statistical errors are estimated by the jackknife method with the 
bin size of 80 trajectories.
We use the all-mode-averaging (AMA) method~\cite{Blum:2012uh,Shintani:2014vja,vonHippel:2016wid} with the deflated Schwartz Alternative Procedure (SAP)~\cite{Luscher:2003qa} and Generalized Conjugate Residual (GCR) \cite{Luscher:2007se}
for the measurements as in our previous work~\cite{Shintani:2018ozy}.
We compute the combination of correlator with high-precision $O^{\rm org}$ and low-precision $O^{\rm approx}$ as
\begin{equation}
  O^{\rm (AMA)}=\frac{1}{N_{\rm org}}\sum^{N_{\rm org}}_{f\in G}\big(O^{{\rm (org)}\,f} - O^{{\rm (approx)}\,f}\big)
 + \frac{1}{N_G}\sum^{N_G}_{g\in G}O^{{\rm (approx)}\,g},
  \label{Eq:ama}
\end{equation}
where the superscript $f,\,g$ denotes the transformation under the lattice symmetry $G$.
In our calculation, it is translational symmetry, {\it e.g.,}
changing the position of the source operator, and 
rotating the temporal direction
using the hypercube symmetry of the configuration.
$N_{\rm org}$ and $N_G$ are the numbers for $O^{\rm org}$
and $O^{\rm approx}$, respectively.
The numbers and the stopping conditions of the quark propagator 
for the high and low-precision
measurements are summarized in Table~\ref{tab:AMA}.
We also take the average of the forward and backward 3-point functions, and
also three 3-point functions with the projector ${\cal P}_{5j}$
in all the three spatial directions $j=x,y,z$ to increase statistics.

In our calculation we obtain $M_\pi = 0.1382(11)$ GeV 
and $M_N = 0.9499(27)$ GeV, which agree
with the previous calculation done with the same 
configuration~\cite{Ishikawa:2018jee,Ishikawa:2019qwn}.
The result of $M_N$ is shown by the solid lines in Fig.~\ref{fig:eff_N},
which is obtained from a single exponential fit of $C_{LS}(t;{\bf 0})$
in the asymptotic region.
Although a little finite volume effect of 3 MeV was observed in the pion mass $M_\pi$ by
comparing with results obtained on the $L=64$ and $L=128$ lattice volumes~\cite{Ishikawa:2018jee},
the finite volume effect was not seen in the nucleon mass $M_N$~\cite{Ishikawa:2019qwn}.
For the renormalization factors, we adopt $Z_V = 1/{\cal R}^t_{V_t}(t;{\bf 0})$ 
with $G_E(0) = 1$, and $Z_A = 0.9650(68)$ evaluated by
the Schr\"odinger functional scheme~\cite{Ishikawa:2015fzw}.
The physical quantities obtained from the DFF method
are compared with the standard analysis of the form factors evaluated
with the same configuration, and also the ones from the larger volume
calculation of (10.9 fm)$^3$ in Ref.~\cite{Shintani:2018ozy}.
The parameters for the larger volume data used in the comparison 
are summarized in Table~\ref{tab:phys_para}.

For a study of finite volume effect for physical quantities obtained
from the DFF method, a small test calculation is 
carried out at a heavier pion mass of $M_\pi = 0.51$ GeV using
the $N_f = 2 + 1$ configurations with $a^{-1} = 2.194$ GeV
generated in Ref.~\cite{Yamazaki:2012hi}.
The parameters for this study are tabulated in Table~\ref{tab:mpi0.5_para}.

\begin{table}[!ht]
\caption{Parameters for the AMA technique used in each choice of 
the source-sink separation ($t_{\rm sep}$): 
the stopping conditions of the quark propagator in the high- and low-precision
measurements ($\epsilon_{\rm high}$ and $\epsilon_{\rm low}$), and 
the number of the measurements for the high- and low-precision calculations ($N_{\rm org}$ and $N_{G}$),
respectively.
\label{tab:AMA}
}
\begin{ruledtabular}
\begin{tabular}{ccccc} 
$t_{\rm sep}$ & $\epsilon_{\rm high}$ & $\epsilon_{\rm low}$ & 
$N_{\rm org}$ & $N_{G}$ \\ \hline
12 & $10^{-8}$ & 0.005 & 4 & 256 \\
14 & $10^{-8}$ & 0.005 & 4 & 1024 \\
16 & $10^{-8}$ & 0.002 & 4 & 2048 \\
\end{tabular}
\end{ruledtabular} 
\end{table}

\begin{table}[!ht]
\caption{
Summary of simulation parameters used in this calculation together
with those in the previous calculation performed on the larger volume~\cite{Shintani:2018ozy}:
the spatial and temporal extents ($L$ and $T$), smearing parameters of the quark field ($A$ and $B$) defined 
in Eq.~(\ref{eq:smeared_quark}), separation of time slice between source and sink operators ($t_{\rm sep}$), 
the total number of the measurement ($N_{\rm meas}= N_{\rm conf} \times N_{G}$) given 
with the numbers for the configurations ($N_{\rm conf}$) and measurements per configuration ($N_G$), respectively.
Recall that the previous results ($L=128$) are useful to be 
compared with the results obtained from the DFF method in this study.
\label{tab:phys_para}
}
\begin{ruledtabular}
\begin{tabular}{cccccc} 
$L$ & $T$ & $A$ & $B$ & $t_{\rm sep}$ & $N_{\rm meas}$ \\ \hline
 64 &  64 & 1.2 & 0.16 & 12 & 25600 \\
    &     &     &      & 14 & 102400 \\
    &     &     &      & 16 & 204800 \\
128 & 128 & 1.2 & 0.14 & 12 & 5120 \\
    &     &     &      & 14 & 6400 \\
    &     &     &      & 16 & 10218 \\
\end{tabular}
\end{ruledtabular} 
\end{table}

\begin{table}[!ht]
\caption{Details of parameters for the finite volume study 
at a heavier pion mass of $M_\pi = 0.51$ GeV:
the spatial and temporal extents ($L$ and $T$), 
separation of time slice between source and sink operators ($t_{\rm sep}$), 
smearing parameters of the quark field ($A$ and $B$) in Eq.~(\ref{eq:smeared_quark}),
the numbers for the configurations ($N_{\rm conf}$) and 
measurements per configuration ($N_G$), respectively.
\label{tab:mpi0.5_para}
}
\begin{ruledtabular}
\begin{tabular}{ccccccc} 
$L$ & $T$ & $t_{\rm sep}$ & $A$ & $B$ & $N_{\rm conf}$ & $N_{G}$ \\ \hline
32 & 48 & 15 & 0.8 & 0.21 & 41 & 16 \\
64 & 64 & 15 & 0.8 & 0.21 & 33 & 16 \\
\end{tabular}
\end{ruledtabular} 
\end{table}

\section{Results}
\label{sec:result}

In this section the results for the momentum derivatives of the nucleon 2- and 3-point
functions are presented.
Physical quantities obtained from the DFF method are compared with
the ones from the standard analysis on the form factors.
The results for the form factors given by the standard 3-point functions 
are summarized in Appendix~\ref{app:form_factors}.

\subsection{Derivative of the nucleon 2-point function}

\begin{figure}[!ht]
 \centering
\includegraphics*[scale=0.50]{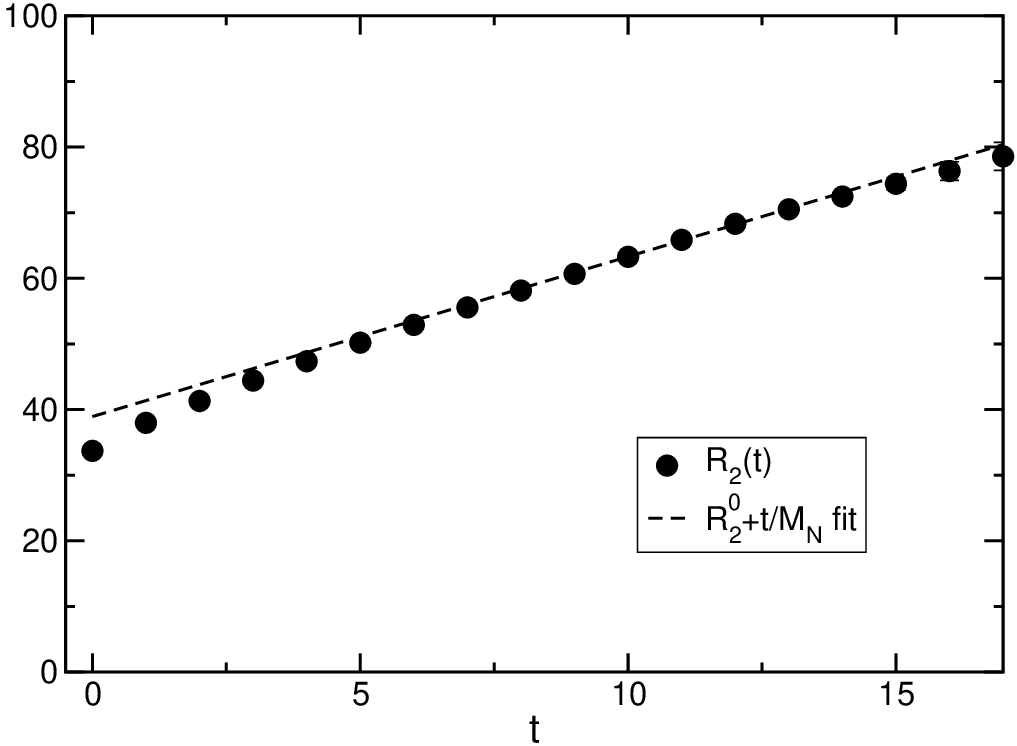}
 \caption{Ratio of the derivative of 2-point function to the standard 2-point function
 as a function of $t$.
 The statistical errors are comparable to the size of the symbols.
 The dashed line represents the linear fit of the large $t$ behavior on $R_2(t)$ as 
 $R_2(t) = R_2^0 + t/M_N$ with the measured $M_N$,
 where $R_2^0$ is only a free parameter.
 \label{fig:der_2pt}
 }
\end{figure}

Figure~\ref{fig:der_2pt} presents the $t$ dependence of the ratio 
$R_2(t)$ defined in Eq.~(\ref{eq:asympt_rat_2pt}), where
the second-order momentum derivative of the 2-point function~(\ref{eq:def_der_2pt}) is divided by the
the zero-momentum 2-point function.
The dashed line represents a linear fit result given with the following form
\begin{equation}
R_2(t) = R_2^0 + \frac{t}{M_N}
\label{eq:fit_R_2}
\end{equation}
with a fit range of $t = 10$--14,
using the linear fixed slope with the measured $M_N$ obtained
from the standard nucleon 2-point function.
This fit result describes the data well in the large $t$ region, though in the small $t$ region 
we observe a deviation from the linear behavior which indicates
the unwanted excited-state contributions appearing in the ratio.
In the following analyses, the fit result $R_2^0$ is utilized to 
eliminate the common contribution in
$R_{{\cal O}_\alpha, (l)}^{k,(n)}(t)$ of Eq.~(\ref{eq:asympt_rat_3pt})
for $n\ge2$, so as to extract the physical quantities of interest
as discussed in Sec.~\ref{sec:der_corr}.

\subsection{MS charge radius $\langle r_E^2 \rangle$}

\begin{figure}[!ht]
 \centering
 \includegraphics*[scale=0.50]{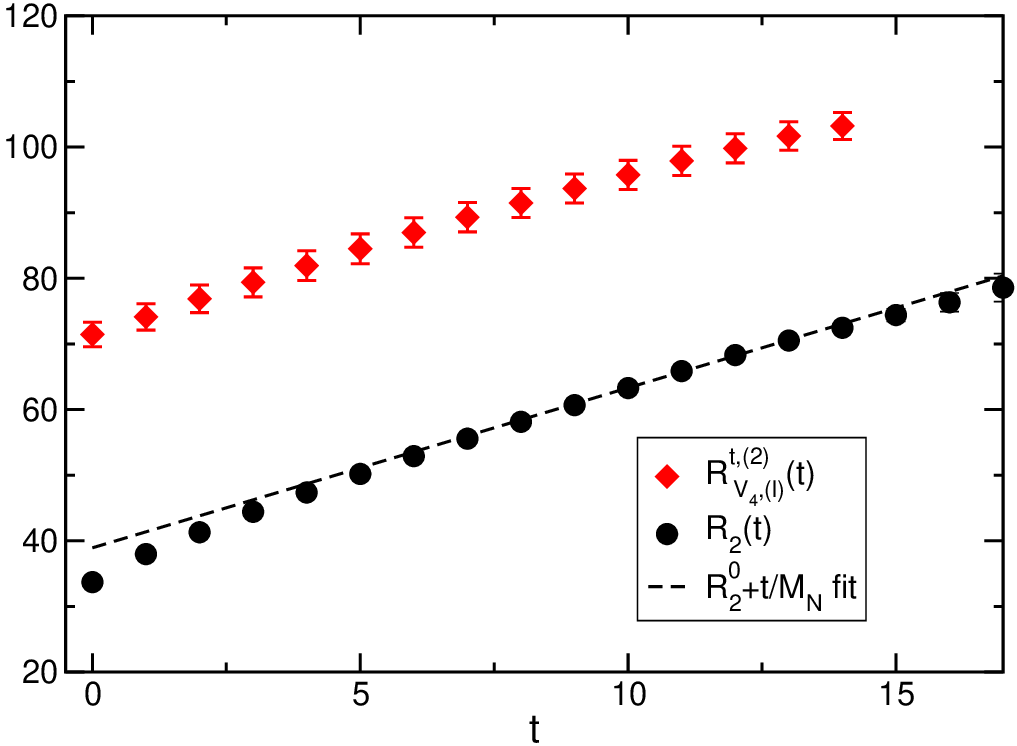}
 \caption{Ratio of the derivative of 3-point function 
 to the standard zero-momentum 3-point function, $R_{V_4,(l)}^{t,(2)}(t)$,
 for $G_E(q^2)$ as a function of $t$
 together with the corresponding one of 2-point function, $R_2(t)$, already 
 shown in Fig.~\ref{fig:der_2pt}.
 A linear fit result of $R_2(t)$ is also plotted as the dashed line.
  \label{fig:der_3pt_GE}
 }
\end{figure}

In the DFF method,
the MS charge radius is extracted from $R_{V_4,(l)}^{t,(2)}(t)$
with the second derivative of the 3-point function $C_{V_4,(l)}^{t,(2)}(t)$,
using its asymptotic form obtained with Eq.~(\ref{eq:der_v4_2}).
The $t$ dependence of $R_{V_4,(l)}^{t,(2)}(t)$ 
is presented in Fig.~\ref{fig:der_3pt_GE} together with the data of $R_2(t)$.
The slope of $R_{V_4,(l)}^{t,(2)}(t)$ reasonably agrees with 
the one of $R_2(t)$ as expected.
At a glance, it is observed that
$R_{V_4,(l)}^{t,(2)}(t)$ has the smaller effect
from excited states than the one appearing in $R_2(t)$,
since the $t$ dependence of $R_{V_4,(l)}^{t,(2)}(t)$ 
exhibits almost linear behavior even in a small $t$ region. 

Considering the second derivative of 
$C^{t}_{V_t}(t;{\bf p})$,
one finds the asymptotic form of $R_{V_4,(l)}^{t,(2)}(t)$ as
\begin{equation}
R_{V_4,(l)}^{t,(2)}(t) = \frac{\langle r_E^2 \rangle}{3} + A + \frac{t}{M_N} ,
\label{eq:fit_R_V4}
\end{equation}
based on the asymptotic form of $C^{t}_{V_t}(t;{\bf p})$ given in Eq.~(\ref{eq:C_GE}), where we use the following relation
\begin{equation}
\left.\frac{\partial^2 f(q^2)}{\partial p_l^2}
\right|_{{\bf p} = 0} = 2 \frac{\partial f(0)}{\partial q^2} .
\end{equation}
with $q^2 = 2M_N(E_N(p^2)-M_N)$ and the condition of ${\bf q}={\bf p}$.

As discussed in Sec~\ref{sec:der_corr},
the last two terms can be removed with $R_2(t)$
or a set of two quantities: $R_2^0$ obtained from $R_2(t)$
(through Eq.~(\ref{eq:fit_R_2})) and the measured $M_N$ obtained from the standard spectroscopy.

\begin{figure}[!ht]
 \centering
 \includegraphics*[scale=0.50]{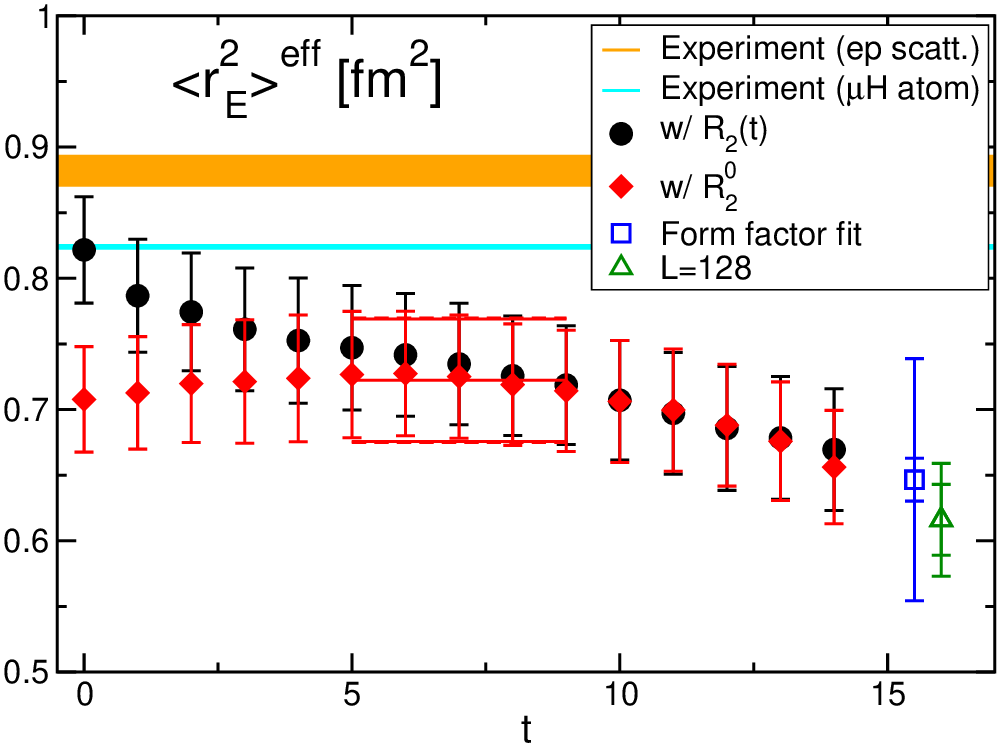}
 \caption{Effective MS charge radius $\langle r_E^2 \rangle^{\rm eff}$
 as a function of $t$, determined 
from $R_{V_4,(l)}^{t,(2)}(t)$ with the data of $R_2(t)$ (circle symbols) 
and the values of $R_2^0$ and $M_N$ (diamond symbols), respectively.
The solid red line represents a constant fit result on the diamond symbols, and 
its statistical error. The dashed lines appearing slightly outside the solid lines
represent the total error.
The MS radius determined from the standard analysis of $G_E(q^2)$ in this calculation
on the $L=64$ lattice volume is plotted by the square symbol, and the one obtained 
on the $L=128$ lattice volume~\cite{Shintani:2018ozy} is also included by the triangle symbol.
The inner and outer errors of these results represent the statistical and total errors, 
respectively.
Here, the total errors are evaluated by adding the statistical and systematic
errors in quadrature. The two horizontal bands represent 
experimental results from $ep$ scattering (upper)~\cite{Mohr:2015ccw} and 
$\mu$-H atom spectroscopy (lower)~\cite{Antognini:1900ns}.
  \label{fig:der_rE}
 }
\end{figure}

Figure~\ref{fig:der_rE} shows the result of the effective
MS charge radius $\langle r_E^2 \rangle$
determined from 
\begin{equation}
\langle r_E^2 \rangle^{\rm eff} = 3\left(R_{V_4,(l)}^{t,(2)}(t)-R_2^0 - \frac{t}{M_N}\right)
\end{equation}
in each $t$ using two measured quantities of $R_2^0$ and $M_N$ (denoted as diamond symbols).
For comparison, we also plot the result (denoted as circle symbols) given by
a naive determination of $\langle r_E^2 \rangle^{\rm eff}$ 
as $3(R_{V_4,(l)}^{t,(2)}(t) - R_2(t))$ using the raw data of $R_2(t)$.

A little $t$ dependence is observed in the naive subtraction due to
the non-negligible excited state contamination in $R_2(t)$ as explained earlier.
Since the former value of $\langle r_E^2 \rangle^{\rm eff}$ exhibits a flat region,
we reliably determine $\langle r_E^2 \rangle$ by a constant fit with the former value
in the region of $t = 5$--9 as drawn by the solid line together with
the statistical error band.
A systematic error is simply estimated by the maximum difference of the raw data in the fit region from the fit result.
The combined error, where the statistical and systematic errors are
added in quadrature, is shown in the figure by the dashed lines,
although they are almost overlapped with the solid lines.
The value of $\langle r_E^2 \rangle$ obtained from the above analysis (denoted as the DFF method)
is tabulated in Table~\ref{tab:results_der}.
In the following analysis, the systematic error of quantities obtained  
from the DFF method will be quoted in the same manner as described above.

The result of $\langle r_E^2 \rangle$ is compared with the one determined from
the fitting of the $q^2$ dependence of $G_E(q^2)$ in Fig.~\ref{fig:der_rE}.
The result of $\langle r_E^2 \rangle$ obtained from the dipole fit defined in Eq.~(\ref{eq:dipole_form})
on the data of $G_E(q^2)$ has 
the smaller statistical error (inner error bar) than that of the above mentioned DFF method, while 
the rather large systematic uncertainty (outer error bar) on the fit result of $G_E(q^2)$
is inevitable due to the choice of the fit form. 
A systematic error is estimated by maximum discrepancy of the results obtained
with different fit forms, namely the quadratic and z-expansion forms as defined 
in Eqs.~(\ref{eq:quad_form}) and (\ref{eq:z-exp_form}). 
Therefore, the results obtained from both the standard and DFF methods, 
are mutually consistent within the total errors whose sizes are comparable.
The two results obtained in this study also reasonably agree with the previous result obtained 
from the standard method in our calculation performed 
on the larger volume ($L=128$) at the physical point~\cite{Shintani:2018ozy}.

\begin{figure}[!ht]
 \centering
 \includegraphics*[scale=0.50]{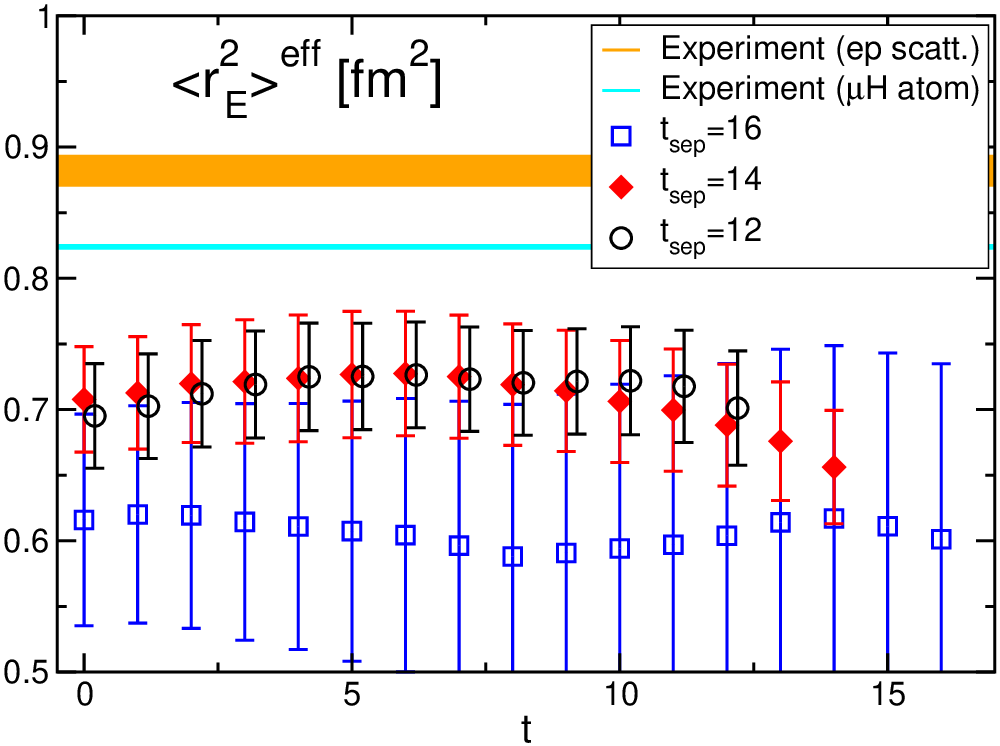}
 \caption{
Comparison of the effective MS charge radius $\langle r_E^2 \rangle^{\rm eff}$ obtained with
$t_{\rm sep}= 12$ (circle symbols), 14 (diamond symbols), and 16 (square symbols), respectively.
The data of $t_{\rm sep}=12$ are slightly shifted to the positive 
$x$ direction for clarity.
The two horizontal bands represent 
experimental results from $ep$ scattering (upper)~\cite{Mohr:2015ccw} and 
$\mu$-H atom spectroscopy (lower)~\cite{Antognini:1900ns}.
\label{fig:der_rE_tsep_dep}
 }
\end{figure}

It is worth mentioning that, compared to the values obtained from the standard calculation
with the form factor $G_E(q^2)$,
the result of $\langle r_E^2 \rangle$ in the DFF method is likely
closer to both experimental values from the electron-proton 
scattering~\cite{Mohr:2015ccw}, 0.882(11) fm$^2$, 
and muonic hydrogen spectroscopy~\cite{Antognini:1900ns}, 0.823(2) fm$^2$,
while there remains a discrepancy of more than 10\%.
It might not be attributed to excited state contributions. 
The reason is that the data given with $t_{\rm sep}=12$ completely agrees with
those with $t_{\rm sep}=14$ as shown in Fig.~\ref{fig:der_rE_tsep_dep} in the DFF method.
The data given with $t_{\rm sep}=16$ is also included in the figure,
though it is too difficult to determine that there is an obvious dependency
with respect to $t_{\rm sep}$ because of its large statistical errors.
In Ref.~\cite{Hasan:2017wwt}, 
a large $t_{\rm sep}$ dependence was reported for the calculation of $\langle r_E^2 \rangle$ using 
another derivative method. We, however, consider that it could be 
caused by their choice of smearing parameters for the quark operators, since our smearing parameters
are highly optimized to eliminate the excited state contributions in the nucleon 2-point and 3-point 
functions. 

\begin{figure}[!ht]
 \centering
 \includegraphics*[scale=0.5]{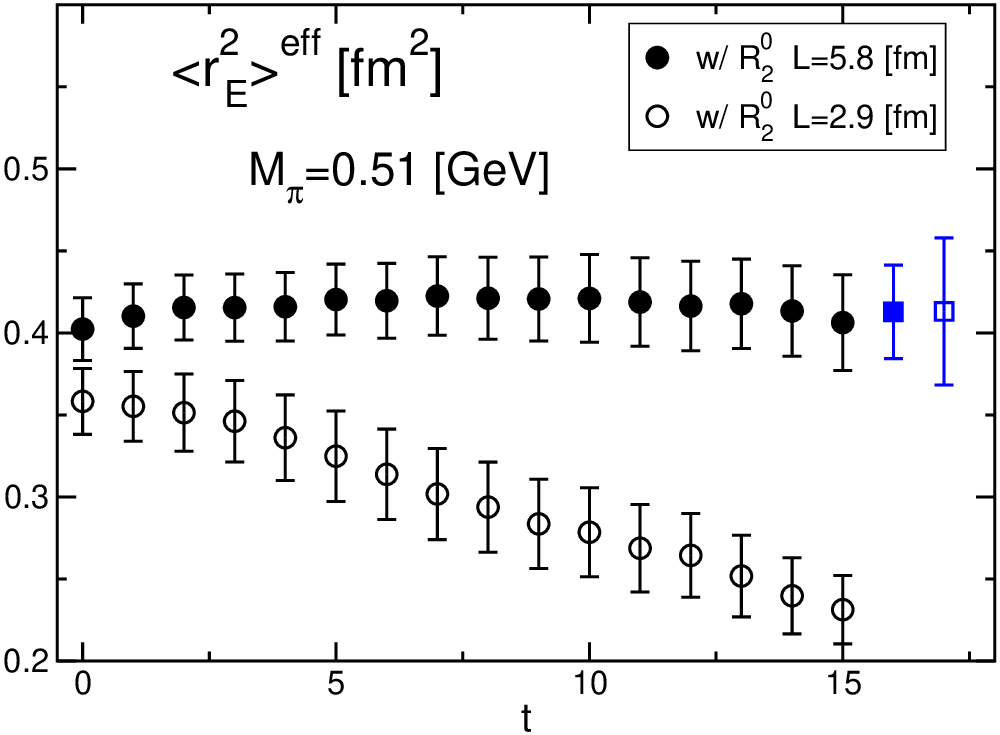}
 \caption{
Finite volume study of $\langle r_E^2 \rangle^{\rm eff}$ at a heavier pion mass of $M_\pi = 0.51$ GeV.
Filled and open circle symbols 
represent the data of the spatial extent of 5.8 fm and 2.9 fm, respectively.
The square symbols correspond to the dipole fit results of the form factor $G_E(q^2)$ 
on the larger (filled symbols) and smaller (open symbols) spatial volume.
  \label{fig:der_rE_m0.5}
 }
\end{figure}

The finite volume effect is other possible source of the systematic errors in the DFF method.
The large effect on the quantity of $\langle r_E^2 \rangle$ is not observed in our standard
analyses with the form factor $G_E(q^2)$ obtained from the $L=64$ and $L=128$ lattice data
shown in Appendix~\ref{app:form_factors}.
A significant effect, however, is reported in a previous study in momentum derivative calculations
of meson correlation functions~\cite{Lellouch:1994zu},
and also found in our pilot study of $\langle r_E^2 \rangle^{\rm eff}$
at a heavier pion mass of $M_\pi = 0.51$ GeV.
Figure~\ref{fig:der_rE_m0.5} shows that
there is a large discrepancy of $\langle r_E^2 \rangle^{\rm eff}$
on volumes of the spacial extent of 2.9 and 5.8 fm at $M_\pi = 0.51$ GeV, while 
the two results determined from the form factor in the standard way are highly consistent with each other
and also the larger volume result from the DFF method.

Thus, for a precise determination of $\langle r_E^2 \rangle$, 
it is an important future task to investigate the systematic uncertainty 
in the DFF method due to the finite size effect near the physical point, 
using our large volume configuration of $L=128$.
Finally, we remark on an attempt of the improved analysis in the DDF method 
for the case of meson form factors~\cite{Lellouch:1994zu,Feng:2019geu}
in order to reduce the finite volume effect. This improvement 
can be also extended to the nucleon form factors.

\begin{table}[!ht]
\caption{
Summary of physical quantities obtained from the DFF method and also the corresponding 
ones evaluated in the standard analysis with the respective form factors. 
The raw data of the form factors are summarized in Appendix~\ref{app:form_factors}.
For comparison, the larger volume ($L=128$) results, which are given by the average of results from
$t_{\rm sep}=12$, 14 and 16 within the standard analysis in our previous calculation~\cite{Shintani:2018ozy},
are also tabulated.
\label{tab:results_der}
}
\begin{ruledtabular}
\begin{tabular}{ccccc}
& $\langle r_E^2 \rangle$ [fm$^2$] & $\mu$ &
$\langle r_M^2 \rangle$ [fm$^2$] & $\langle r_A^2 \rangle$ [fm$^2$] \\
\hline
DFF method & 0.722(47)(8) & 4.337(73)(31) & 0.397(46)(24) & 0.341(35)(13) \\
form factor($L=64$)  & 0.646(16)(91) & 4.36(4)(82) & 0.58(2)(2.67) & 0.308(17)(52)\\
form factor($L=128$) & 0.616(27)(33) & 4.41(14)(33) & 0.500(51)(440) & 0.283(30)(77) \\
\end{tabular}
\end{ruledtabular} 
\end{table}

\subsection{Magnetic moment $\mu$}

The magnetic moment $\mu = G_M(0)$ in the DFF method
is calculated from $R^{5j,(1)}_{V_i,(k)}(t)$ with
the first derivative of the 3-point function $C^{5j,(1)}_{V_i,(k)}(t)$, 
using its asymptotic form obtained with Eq.~(\ref{eq:der_vi_1}).
The effective magnetic moment is defined by
\begin{equation}
\mu^{\rm eff} = 2 M_N R^{5j,(1)}_{V_i,(k)}(t) 
\label{eq:def_eff_mu}
\end{equation}
with $k\ne i\ne j$, whose form can be read off by considering the derivative of the asymptotic form of
$C^{5j}_{V_i}(t;{\bf p})$ in Eq.~(\ref{eq:C_GM}).
It is here worth pointing out that the ratio $R^{5j,(1)}_{V_i,(k)}(t)$ is supposed to show no dependence of $t$
in the asymptotic region.

\begin{figure}[!ht]
 \centering
 \includegraphics*[scale=0.50]{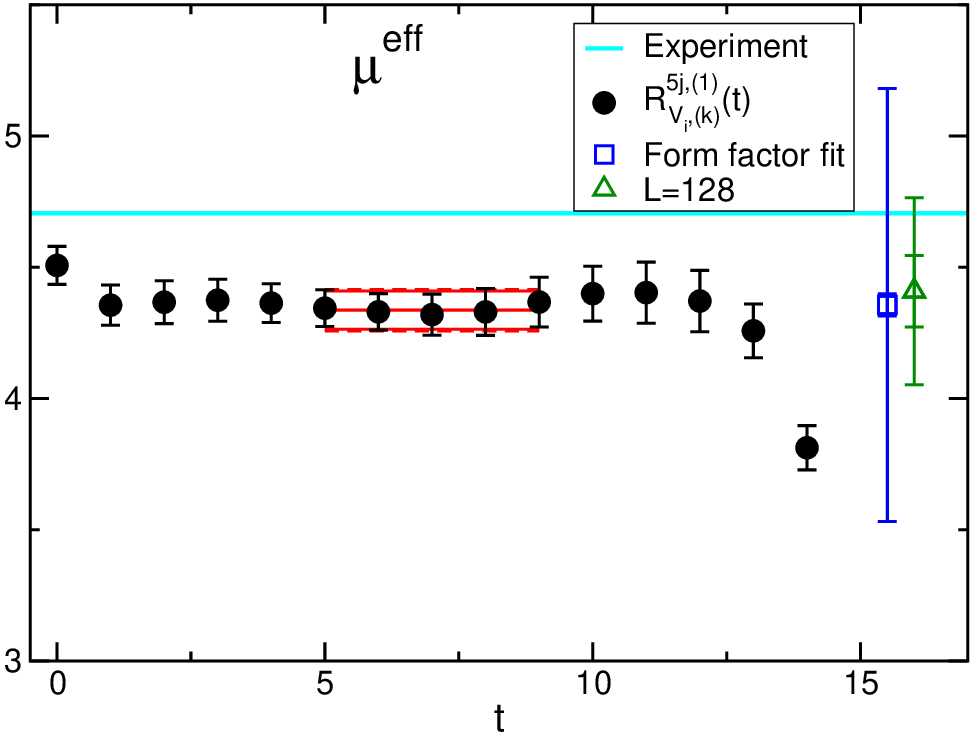}
 \caption{Same as Fig.~\ref{fig:der_rE} for the effective magnetic moment $\mu^{\rm eff}$ defined
in Eq.~(\ref{eq:def_eff_mu}).
The cyan band represents the experimental result~\cite{Zyla:2020zbs}.
  \label{fig:der_mu}
 }
\end{figure}

Figure~\ref{fig:der_mu} presents that
the data of $\mu^{\rm eff}$ exhibits a long flat region.
The value of $\mu$ is again determined from a constant fit of the data
in the region of $t = 5$--9.
The fit result of $\mu$ (denoted as the DFF method) is tabulated in Table~\ref{tab:results_der}.
As a result, the total error of the DFF result is dominated only
by the statistical error as same in the case of $\langle r_E^2 \rangle^{\rm eff}$.
As shown in Fig.~\ref{fig:der_mu}, 
a reasonable consistency between $\mu$ obtained by the DFF method
and the dipole fit result of $G_M(q^2)$ is observed.
However, the fit result of $G_M(q^2)$ receives the large systematic 
uncertainty due to the choice of the fit form.
Indeed, it is mainly caused by the z-expansion fit result of $G_M(q^2)$, where 
unnatural vending down behavior towards $q^2=0$ is observed 
in the data of the $L=64$ lattice volume.
It was less likely to happen in our previous study with the larger volume ($L=128$), 
where the lowest $q^2$ becomes closer to $q^2=0$.

Recall that for the case of $G_M(q^2)$, $G_M(0)$ cannot be directly measured 
in the standard method for kinematical reasons. Therefore, the
determination by the fitting of the $q^2$ dependence of $G_M(q^2)$
is sometimes suffered from the large model dependence of the fit form 
because of the absence of the $G_M(0)$ data. For this reason, the systematic error 
on the DFF result in the smaller volume calculation 
is in general much better controlled than that of the fit result obtained even 
from the larger volume calculation.

\begin{figure}[!ht]
 \centering
 \includegraphics*[scale=0.50]{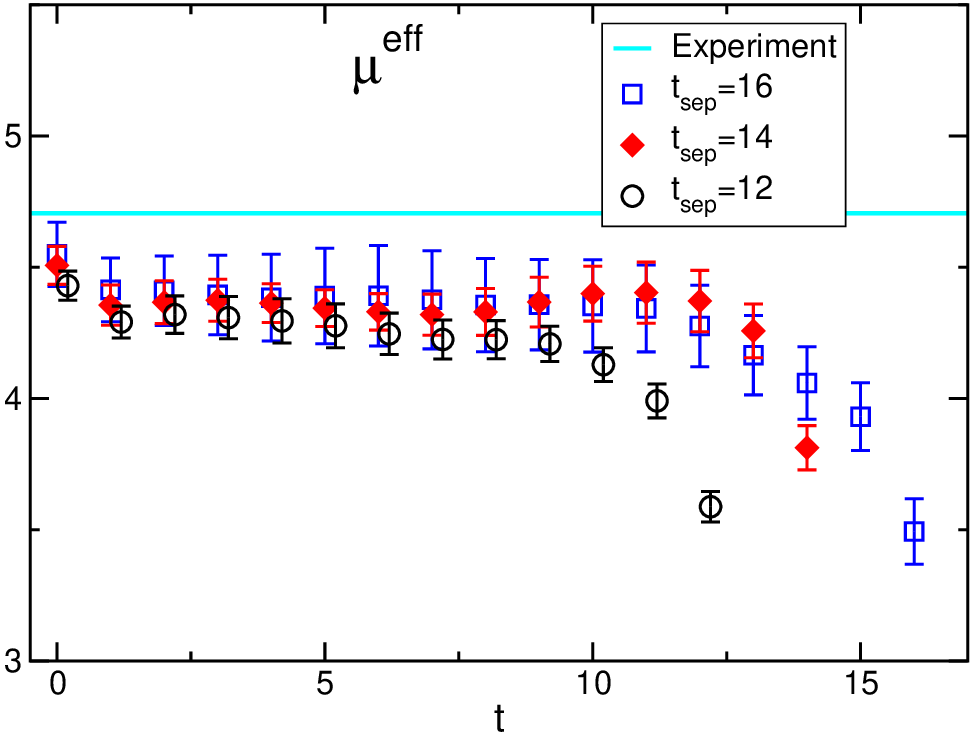}
 \caption{Same as Fig.~\ref{fig:der_rE_tsep_dep} for 
the effective magnetic moment $\mu^{\rm eff}$. 
The cyan band represents the experimental result~\cite{Zyla:2020zbs}.
  \label{fig:der_mu_tsep_dep}
 }
\end{figure}

The result in the DFF method is in good agreement with
the two fit results of $G_M(q^2)$ obtained in this study ($L=64$) and also from 
the larger volume calculation ($L=128$)~\cite{Shintani:2018ozy}. 
However, the significant reduction of the systematic error in the DFF method
may expose an underestimation of $\mu$ in comparison with the experimental value, $\mu = 4.70589$,
in PDG20~\cite{Zyla:2020zbs}.
In order to investigate possible other systematic errors in the DFF method, 
the direct comparison of the data with $t_{\rm sep}=12$, 14, and 16 is first presented
in Fig.~\ref{fig:der_mu_tsep_dep}.
The three data statistically agree with each other in the middle $t$ region. 
We thus consider that the present data set shows no significant uncertainty  
associated with the excited state contamination
in the data of $t_{\rm sep}=14$, though 
the statistical error of $t_{\rm sep}=16$ is larger than
those for the other two data.

\begin{figure}[!ht]
 \centering
 \includegraphics*[scale=0.50]{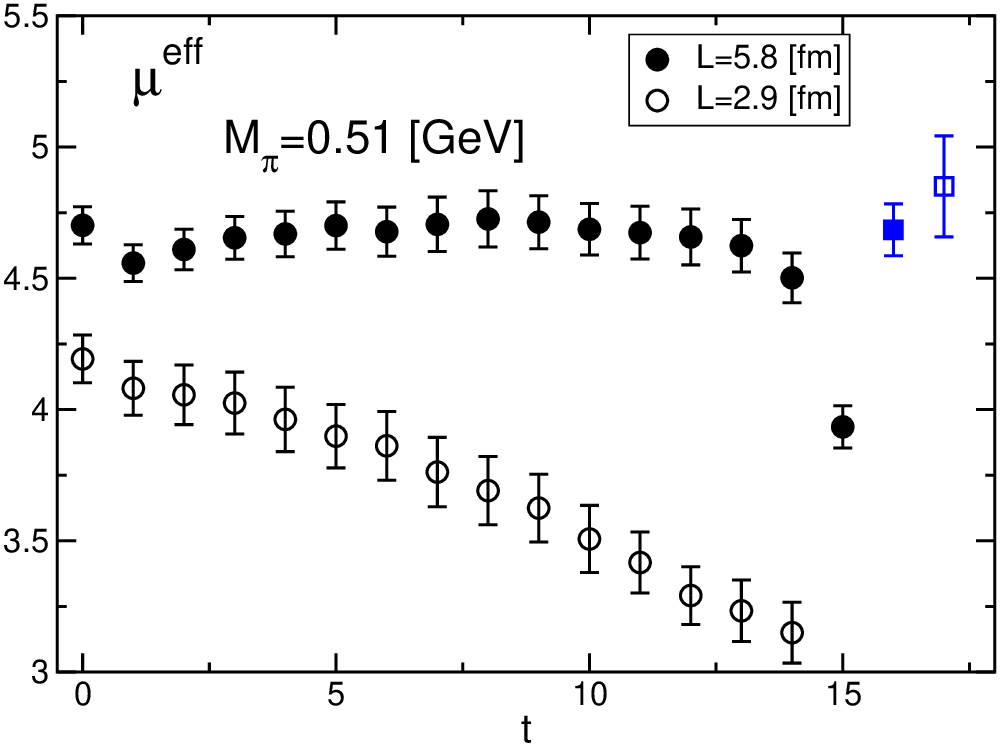}
 \caption{Same as Fig.~\ref{fig:der_rE_m0.5} for the effective magnetic moment $\mu^{\rm eff}$.
  \label{fig:der_mu_m0.5}
 }
\end{figure}

We rather consider that the discrepancy from the experiment might be 
caused again by a finite volume effect. As shown in Fig.~\ref{fig:der_mu_m0.5}, 
in our pilot study of the DFF method at $M_\pi = 0.51$ GeV, we observe a large 
finite volume effect of more than 10\% in comparison with the data on the (2.9 fm)$^3$ 
and (5.8 fm)$^3$ volumes even far away from the physical point.
Note that the two dipole-fit results obtained with two spatial extents of $L=64$ and $L=128$
at the physical point agree with each other within the standard method, and thus such a large effect is not detected in the standard form factor calculation.
More detailed investigations are required for full control of possible systematic errors 
in the DFF method in order to resolve the discrepancy mentioned above.

\subsection{MS magnetic radius $\langle r_M^2 \rangle$}

Considering the third derivative of $C^{5j}_{V_i}(t;{\bf p})$ and its asymptotic form obtained with Eq.~(\ref{eq:C_GM}),
$R^{5j,(3)}_{V_i,(kl)}(t)$ yields the following asymptotic form:
\begin{equation}
R^{5j,(3)}_{V_i,(kl)}(t) = \frac{\mu}{2M_N}
\left( \frac{\langle r_M^2 \rangle}{3} + \frac{1}{2 M_N^2} + A + \frac{t}{M_N}
\right) ,
\label{eq:fit_R_A}
\end{equation}
with $l\ne k$ where $l$ and $k$ are the directions of the derivative
defined in Eq.~(\ref{eq:der_vi_3}).
When $l = k$ is set in Eq.~(\ref{eq:der_vi_3}),
the right hand side of the above equation is multiplied by a factor of three.
In our analysis all the $l$ data are averaged.

\begin{figure}[!ht]
 \centering
 \includegraphics*[scale=0.50]{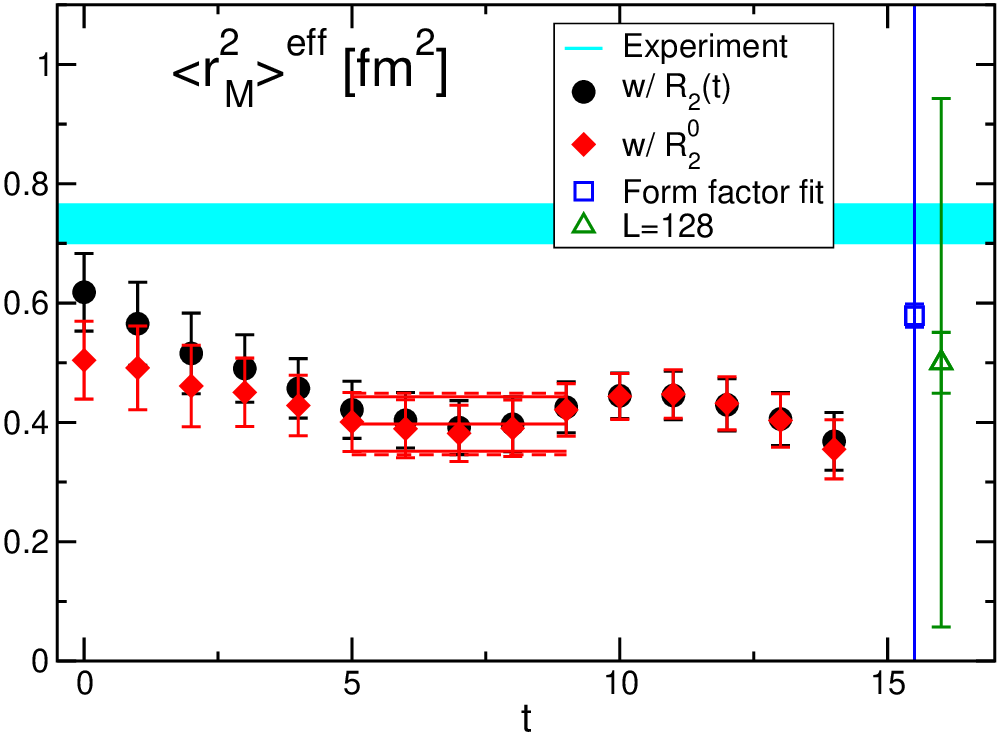}
 \caption{Same as Fig.~\ref{fig:der_rE} for the effective MS magnetic radius 
$\langle r_M^2 \rangle^{\rm eff}$ obtained from $R^{5j,(3)}_{V_i,(kl)}(t)$.
The cyan band represents the experimental result~\cite{Zyla:2020zbs}.
  \label{fig:der_rM}
 }
\end{figure}

Using the asymptotic form of Eq.~(\ref{eq:fit_R_A}),
the effective MS magnetic radius in the DFF method is determined by
\begin{equation}
\langle r_M^2 \rangle^{\rm eff} = 
3\left( \frac{R^{5j,(3)}_{V_i,(kl)}(t)}{R^{5j,(1)}_{V_{i},(k)}(t)}
- \frac{1}{2 M_N^2} - R_2^0 - \frac{t}{M_N}
\right) 
\label{eq:def_eff_rM}
\end{equation}
with the first derivative $R^{5j,(1)}_{V_{i},(k)}(t)$ 
described in the previous subsection.
The data of $\langle r_M^2 \rangle^{\rm eff}$ plotted
in Fig.~\ref{fig:der_rM} exhibits the milder $t$ dependence in small $t$ region
compared to the data from a naive subtraction with the raw data of $R_2(t)$,
$3( {R^{5j,(3)}_{V_i,(kl)}(t)}/{R^{5j,(1)}_{V_i,(k)}(t)}-{1}/{(2 M_N^2)} - R_2(t))$
for the same reason as in the case of $\langle r_E^2 \rangle^{\rm eff}$.
In the middle $t$ region the two results are mutually consistent,
so that we determine the value of $\langle r_M^2 \rangle$ by a constant fit of
$\langle r_M^2 \rangle^{\rm eff}$ in the region of $t=5$--9,
which is plotted by the solid lines in Fig.~\ref{fig:der_rM},
and tabulated in Table~\ref{tab:results_der}.
Although the statistical error of $\langle r_M^2 \rangle$ 
obtained from the fit of $G_M(q^2)$ on the $L=64$ lattice volume is 
smaller than that of the DFF method, 
much larger systematic error regarding the choice of fit function
makes their total accuracy worse than the DFF result. 

\begin{figure}[!ht]
 \centering
 \includegraphics*[scale=0.50]{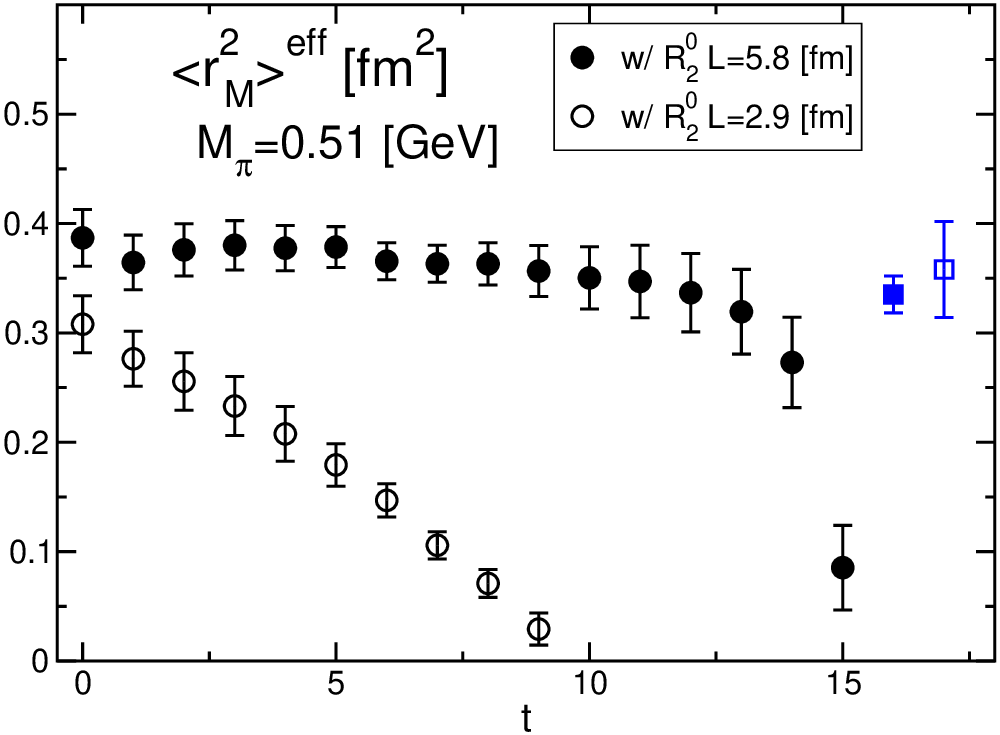}
 \caption{Same as Fig.~\ref{fig:der_rE_m0.5} for the effective MS magnetic radius $\langle r_M^2 \rangle^{\rm eff}$.
  \label{fig:der_rM_m0.5}
 }
\end{figure}

While the result of $\langle r_M^2 \rangle$ given in the DFF method
has smaller systematic error, its central value is much smaller than the experimental value,
$\langle r_M^2 \rangle = 0.733(32)$ fm$^2$~\cite{Zyla:2020zbs}.
It might be caused by a finite volume effect in the DFF calculation,
since this quantity is indeed considered to 
be more sensitive to finite volume effects compared to other quantities 
discussed earlier. 
In the DFF method,
the quantity of $\langle r_M^2 \rangle$ requires the third derivative of
the 3-point function {\it i.e.,} the third moment of the 3-point function in coordinate space.
Therefore, the larger spatial extent is naturally required for its higher moment compared 
to the other quantities considered in the previous subsections.
Figure~\ref{fig:der_rM_m0.5} shows that 
in our pilot calculation of $\langle r_M^2 \rangle^{\rm eff}$ 
at $M_\pi = 0.51$ GeV, more significant finite volume effect is observed
than the case of $\langle r_E^2\rangle^{\rm eff}$ in Fig.~\ref{fig:der_rE_m0.5}:
the data of $\langle r_M^2 \rangle^{\rm eff}$ on the smaller volume
becomes negative in a large $t$ region in contrast to 
that of $\langle r_E^2\rangle^{\rm eff}$.

\begin{figure}[!ht]
 \centering
 \includegraphics*[scale=0.50]{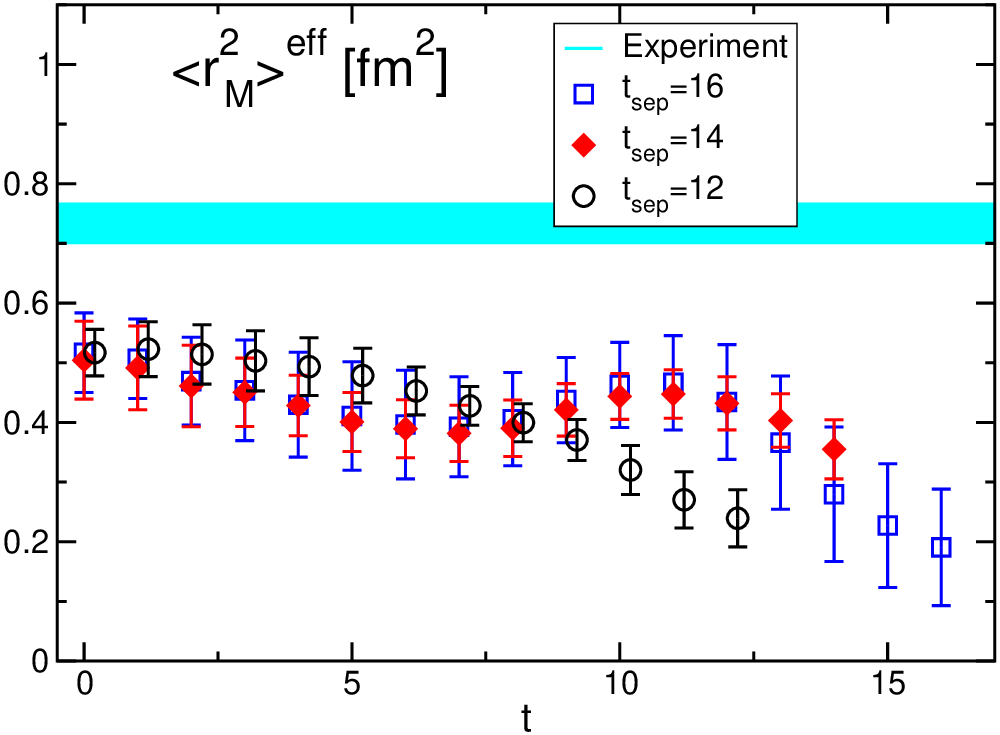}
 \caption{Same as Fig.~\ref{fig:der_rE_tsep_dep} for 
the effective MS magnetic radius $\langle r_M^2 \rangle^{\rm eff}$.
The cyan band represents the experimental result~\cite{Zyla:2020zbs}.
  \label{fig:der_rM_tsep_dep}
 }
\end{figure}

Another possible source of systematic errors comes from the excited state contamination.
Comparing the three data sets using $t_{\rm sep}=12,$ 14, and 16 in 
Fig.~\ref{fig:der_rM_tsep_dep}, the $t_{\rm sep}$ dependence is not clearly
seen due to their large statistical fluctuations.
In a future work, it is important to investigate the above two systematic errors by 
calculations with the larger volume ($L=128$) and the large variation of $t_{\rm sep}$.

\subsection{MS axial radius $\langle r_A^2 \rangle$}

\begin{figure}[!ht]
 \centering
 \includegraphics*[scale=0.50]{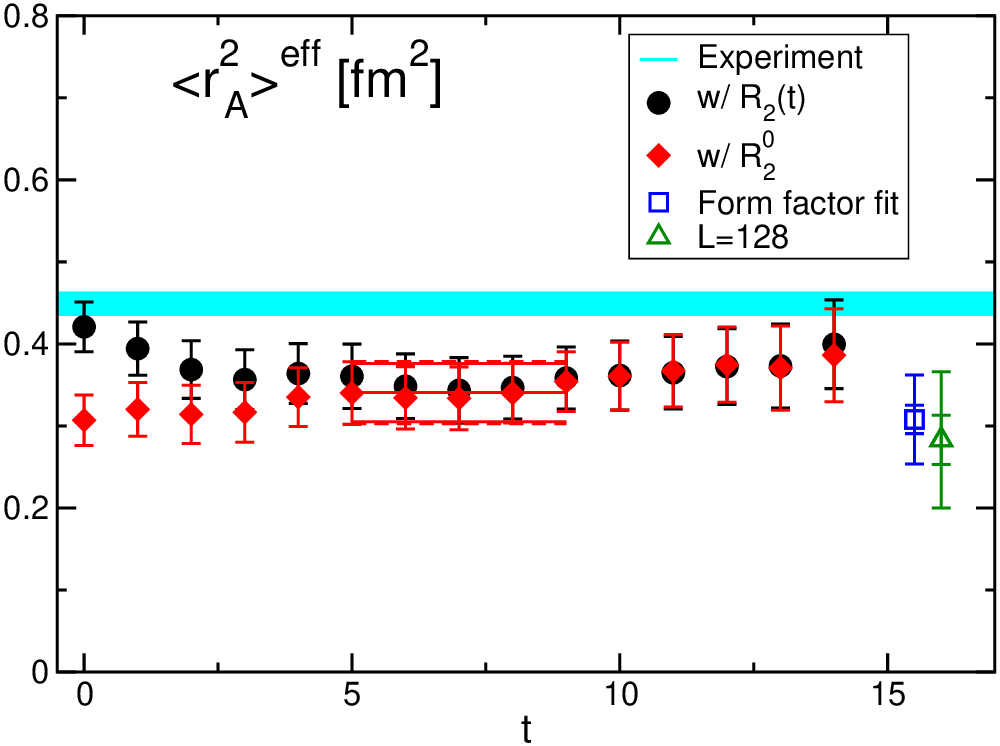}
 \caption{Same as Fig.~\ref{fig:der_rE} for 
the effective MS axial-vector radius $\langle r_A^2 \rangle^{\rm eff}$ obtained from $R_{A_j, (i)}^{5j,(2)}(t)$.
The cyan band represents the experimental result~\cite{Bernard:2001rs,Bodek:2007ym}.
  \label{fig:der_rA}
 }
\end{figure}

For the MS axial radius $\langle r_A^2 \rangle$, 
the same analysis as in the case of $\langle r_E^2 \rangle$ with $R_{V_4,(l)}^{t,(2)}(t)$
is performed for $R_{A_j,(i)}^{5j,(2)}(t)$,
where $i \ne j$ in the subscript expresses the direction of the derivative 
defined in Eq.~(\ref{eq:der_a3_2_i}).
The asymptotic form of $R_{A_j,(i)}^{5j,(2)}(t)$ is given by
\begin{equation}
R_{A_j, (i)}^{5j,(2)}(t) = \frac{\langle r_A^2 \rangle}{3}
+ A + \frac{t}{M_N} .
\end{equation}
The effective MS axial-vector radius is defined by
\begin{equation}
\langle r_A^2 \rangle^{\rm eff} = 3\left(
R_{A_j,(i)}^{5j,(2)}(t) - R_2^0 - \frac{t}{M_N}
\right) ,
\end{equation}
and its value is plotted as a function of $t$ in Fig.~\ref{fig:der_rA}.
The data is compared with the one determined from
a naive subtraction with the raw data of $R_2(t)$ as
$3(R_{A_j,(i)}^{5j,(2)}(t) - R_2(t))$.
Both estimations for $\langle r_A^2 \rangle^{\rm eff}$
exhibit a reasonably flat behavior in the middle $t$ region, respectively.
We determine the value of $\langle r_A^2 \rangle$
by a constant fit of the data of $\langle r_A^2 \rangle^{\rm eff}$
in the region of $t=5$--9.  The result obtained by the DFF method is tabulated
in Table~\ref{tab:results_der}. As shown in Fig.~\ref{fig:der_rA}, 
the DFF result is fairly consistent with the two dipole-fit results of $F_A(q^2)$
obtained in this study ($L=64$) and also from the larger volume calculation ($L=128$)~\cite{Shintani:2018ozy}.
Again, the total accuracy of the dipole fit results is slightly worse than the DFF result that can avoid
any model dependence. 

\begin{figure}[!ht]
 \centering
 \includegraphics*[scale=0.50]{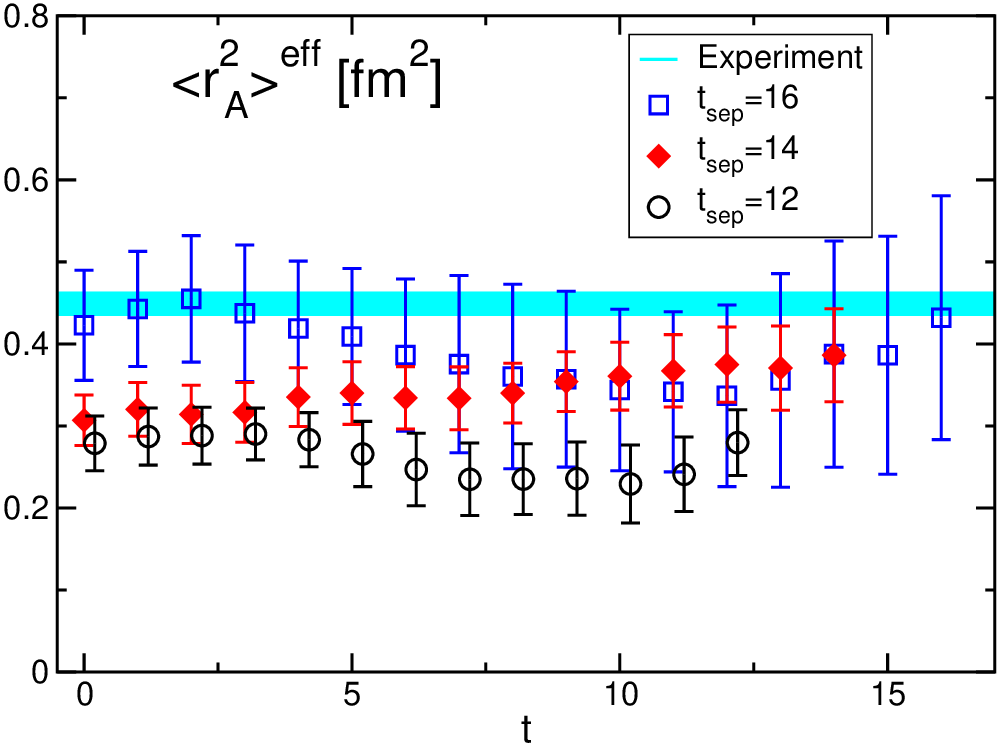}
 \caption{Same as Fig.~\ref{fig:der_rE_tsep_dep} for 
the effective MS axial-vector radius $\langle r_A^2 \rangle^{\rm eff}$.
The cyan band represents the experimental result~\cite{Bernard:2001rs,Bodek:2007ym}.
  \label{fig:der_rA_tsep_dep}
 }
\end{figure}

\begin{figure}[!ht]
 \centering
 \includegraphics*[scale=0.50]{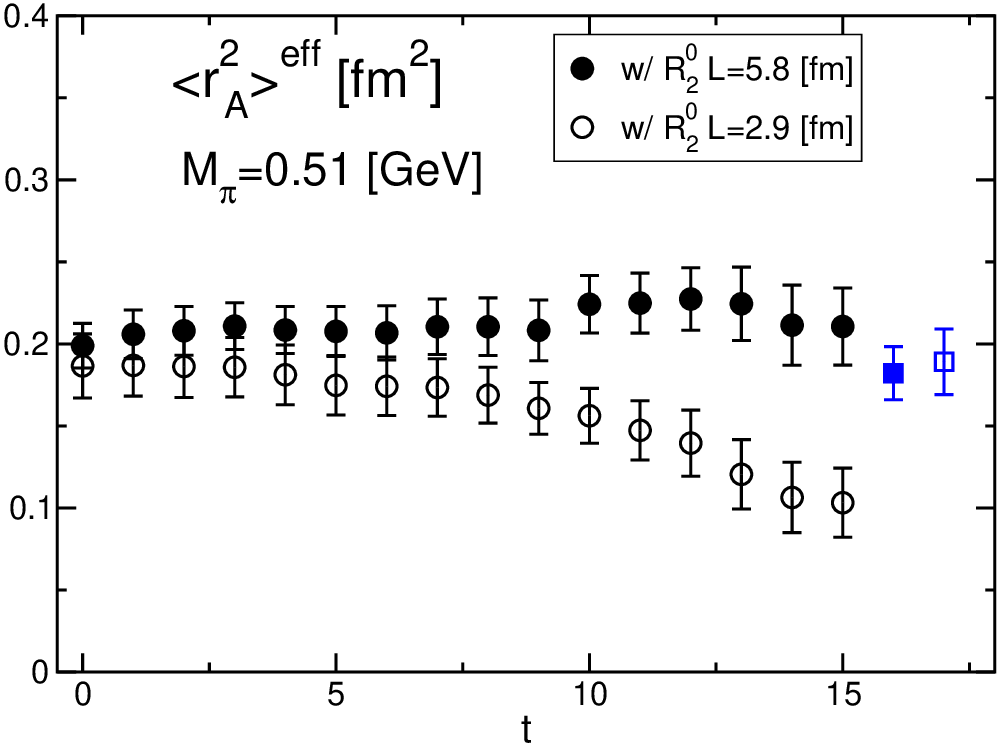}
 \caption{Same as Fig.~\ref{fig:der_rE_m0.5} for the effective MS axial-vector 
 radius $\langle r_A^2 \rangle^{\rm eff}$.
  \label{fig:der_rA_m0.5}
 }
\end{figure}

As in the case of the dipole fit results,
the result of the DFF method is little smaller 
than the experiment~\cite{Bernard:2001rs,Bodek:2007ym}, 0.449(13) fm$^2$.
It can be attributed to excited state contamination.  
It is simply because a visible difference between the data from  
$t_{\rm sep}=12$ and 14 is seen in the flat region of $\langle r_A^2 \rangle^{\rm eff}$
as a function of $t$ as shown in 
Fig.~\ref{fig:der_rA_tsep_dep}.
Furthermore, the data of $t_{\rm sep}=16$ statistically agrees with
the experiment, though the statistical uncertainties are not small enough to
make a firm conclusion.
In addition, recently, it was reported that a large effect due to the excited state contamination
exists in determination of $\langle r_A^2 \rangle$ 
even from the fitting of the $q^2$ dependence of $F_A(q^2)$~\cite{Jang:2019vkm,Park:2021ypf}.
Since we have only a few variations of $t_{\rm sep}$,
we will need to verify whether or not the discrepancy from the experiment can be 
explained by the systematic uncertainties from the excited state contamination 
using the data with the larger variation of $t_{\rm sep}$.
It is worth pointing out that
the systematic error associated with the finite volume effect in
$\langle r_A^2 \rangle^{\rm eff}$ might be smaller than the other quantities obtained 
from the DFF method discussed earlier.
This is expected from our pilot calculation at $M_\pi = 0.51$ GeV
as shown in Fig.~\ref{fig:der_rA_m0.5}.
The difference between the data on the larger and smaller volumes is about 10\%
in the middle $t$ region, which it is much smaller than those of 
other quantities as shown in Figs.~\ref{fig:der_rE_m0.5}, \ref{fig:der_mu_m0.5}, and \ref{fig:der_rM_m0.5}. 

The size of the finite volume effect is supposed to depend on the MS radius of the target
form factor. If the MS radius is small, in other words, the form factor has a broad
shape in the $q^2$ space, its moment in the coordinate space is less
sensitive to the finite volume. This is because the narrow spatial distribution 
in the coordinate space is given by the inverse Fourier transform of the broad form factor in a classical argument.~\footnote{For more details, see, an intuitive argument for the required spatial size in extraction of the MS radius for the spatial distribution
that falls exponentially at large distances as described in Refs.~\cite{{Ishikawa:2018rew}, {Sick:2018fzn}}.} 

Indeed, the three values of $\langle r_A^2 \rangle$ at $M_\pi = 0.51$ GeV, which are obtained from 
the DFF method on the larger volume ($L=64$) and the dipole fit results obtained from both two volumes ($L=32$ and $L=64$) as shown in Fig.~\ref{fig:der_rA_m0.5}, are certainly smaller than those for $\langle r_E^2 \rangle$ 
and $\langle r_M^2 \rangle$. This observation leads to the expectation that $\langle r_A^2 \rangle^{\rm eff}$ has 
the smaller finite volume effect. This expectation should remain valid at the physical $M_\pi$,
since the experimental value of $\langle r_A^2\rangle$ 
is smaller than those of $\langle r_E^2\rangle$ and $\langle r_M^2\rangle$.

\subsection{$F_P(0)$ from the DFF method}

As same in the case of $G_M(0)$, $F_P(0)$ cannot be directly measured 
in the standard method for kinematical reasons.
In the DFF method, the value of $F_P(0)$ is accessible in two ways.
One uses $R_{A_i,(ij)}^{5j,(2)}(t)$ with $i \ne j$ as defined in Eq.~(\ref{eq:der_a3_2_j}), 
while the other uses $R_{A_j,(j)}^{5j,(2)}(t)$ defined in Eq.~(\ref{eq:der_ai_2}).
Their asymptotic forms are given by
\begin{eqnarray}
R_{A_i,(ij)}^{5j,(2)}(t) &=& \frac{F_P(0)}{2M_N g_A},
\label{eq:Fp_R_Ai}\\ 
R_{A_j,(j)}^{5j,(2)}(t) &=& \frac{\langle r_A^2 \rangle}{3}
+ A + \frac{t}{M_N} + \frac{F_P(0)}{M_N g_A} ,
\label{eq:Fp_R_Aj}
\end{eqnarray}
where $i \ne j$. Using them, the effective value of
$2M_N F_P(0)/g_A$ can be defined in two ways as
\begin{eqnarray}
\frac{2M_N F_P^{\rm eff}(0)}{g_A} &=&
4 M_N^2 R_{A_i,(ij)}^{5j,(2)}(t) 
\label{eq:eff_F_P_1}
\\
&=&
2 M_N^2 \left( R_{A_j,(j)}^{5j,(2)}(t) - R_{A_j,(i)}^{5j,(2)}(t)
\right) ,
\label{eq:eff_F_P_2}
\end{eqnarray}
where the second term 
of $R_{A_j,(i)}^{5j,(2)}$ appearing in Eq.~(\ref{eq:eff_F_P_2}) is 
the one used for the determination of $\langle r_A^2 \rangle^{\rm eff}$
as discussed in the last subsection.

\begin{figure}[!ht]
 \centering
 \includegraphics*[scale=0.50]{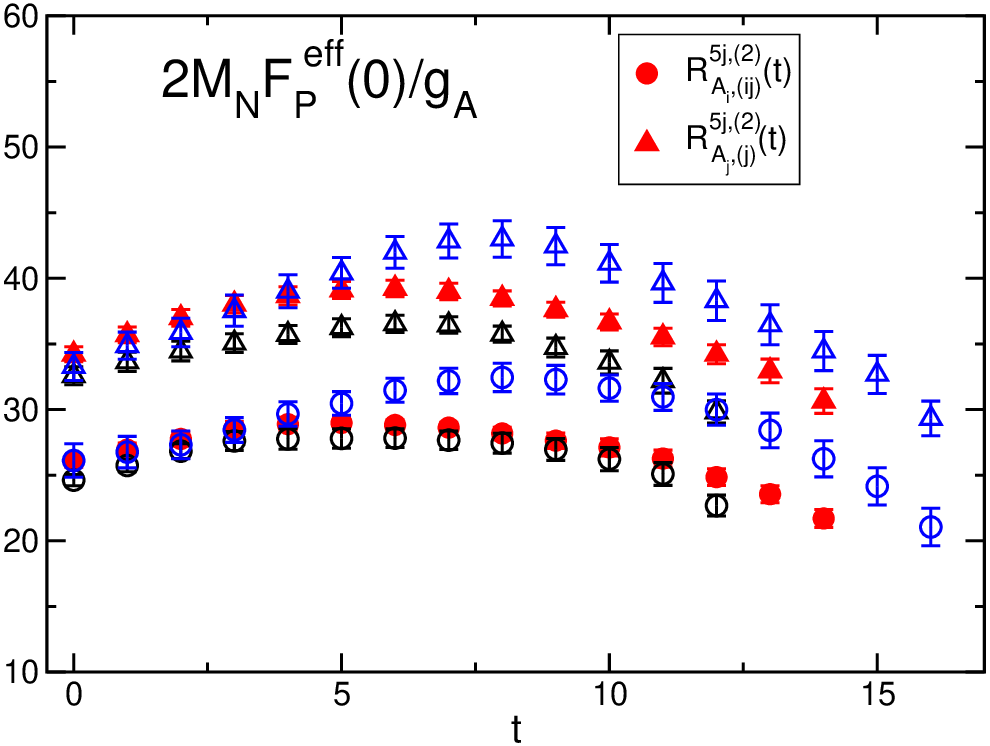}
 \caption{Effective induced pseudoscalar coupling $2 M_N F_P(0)/g_A$ evaluated from 
 $R_{A_i,(ij)}^{5j,(2)}(t)$ (circle symbols) and $R_{A_j,(j)}^{5j,(2)}(t)$ (triangle symbols).
 The black, red and blue symbols represent the data obtained with $t_{\rm sep} = 12, 14$ and 16.
  \label{fig:der_fp_zero}
 }
\end{figure}

Figure~\ref{fig:der_fp_zero} shows that
the two different estimations of $2M_N F_P^{\rm eff}(0)/g_A$ with $t_{\rm sep}=14$
(denoted with filled symbols)
provide inconsistent results: the result obtained from Eq.~(\ref{eq:eff_F_P_1}) is much smaller than 
that of Eq.~(\ref{eq:eff_F_P_2}).
The same behavior is also seen in the data from the smaller and larger source-sink 
separations for $t_{\rm sep}=12$ and 16.
Note that the expected value of $2 M_N F_P(0)/g_A$ is much larger than 
these two observed values according to the following reasons:
(1) $2 M_N F_P(q^2_1)/g_A \sim 40 $ is observed even at the lowest non-zero $q^2 = q_1^2$
in the standard method with $t_{\rm sep}=14$.
(2) The $q^2$ dependence of $F_P(q^2)$ is expected to rapidly increase 
in the limit of $q^2 \to 0$ as shown in Appendix~\ref{app:form_factors}.

\begin{figure}[!ht]
 \centering
 \includegraphics*[scale=0.5]{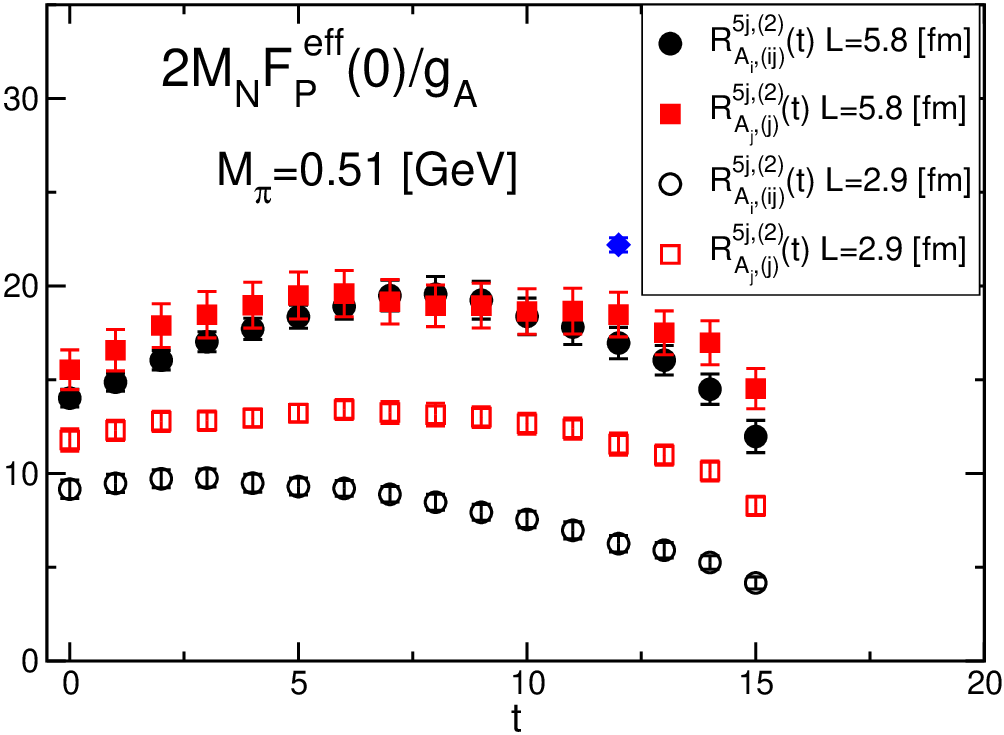}
 \caption{
Finite volume study of effective induced pseudoscalar coupling $2 M_N F_P(0)/g_A$ at 
a heavier pion mass of $M_\pi = 0.51$ GeV.
Filled and open circle symbols represent the data of the spatial extent of 5.8 fm and 2.9 fm, respectively.
Circle and square symbols are determined with 
the definitions given in Eqs.~(\ref{eq:eff_F_P_1}) and (\ref{eq:eff_F_P_2}), respectively.
The single diamond symbol represents the value of $2M_N F_P(0)/g_A$ evaluated by examining
the $q^2$ dependence of the form factor $F_P(q^2)$ on the larger volume ($L=128$).
  \label{fig:der_fp_zero_m0.5}
 }
\end{figure}

This difference between the two estimations of $2M_N F_P^{\rm eff}(0)/g_A$ 
can be attributed to a finite volume effect.
This is simply because a similar trend is observed 
in our pilot calculation at $M_\pi = 0.51$ GeV, where
the discrepancy between the two results becomes resolved in 
a larger volume calculation as shown in Fig.~\ref{fig:der_fp_zero_m0.5}.
A large systematic effect stemming from the finite volume is 
easily understood in this case, since 
the induced pseudoscalar form factor in the coordinate
space, which corresponds to the one given by the inverse Fourier transform of $F_P(q^2)$,
has a very broad structure.
This is naively expected from the fact that $F_P(q^2)$ 
has a sharp peak near the origin at the physical $M_\pi$ corresponding to 
the large contribution of the pion pole in the pion pole dominance model.
The moment of such a broad function in the coordinate space could not
avoid the strong dependence of the finite volume.

In addition to the finite volume effect, in our previous studies,
we observed other problem that the lattice data of $F_P(q^2)$ differs from
the pion pole dominance model~\cite{Sasaki:2007gw,Ishikawa:2018rew},
which can be explained by large excited state 
contamination~\cite{Shintani:2018ozy}.
Similar discussions were reported 
in Refs.~\cite{Bar:2018xyi,Jang:2019vkm,Alexandrou:2020okk,Jang:2019jkn,Park:2021ypf}.
In order to fully resolve the problems, 
more comprehensive investigations are necessary in this particular quantity.

\section{Summary}
\label{sec:summary}

We have calculated the MS radii and magnetic moment 
for the isovector nucleon form factors using the DFF method, 
which is a direct calculation method of the derivative
of the form factor proposed in Ref.~\cite{Aglietti:1994nx}, 
in the $N_f = 2+1$ QCD near the physical point on the (5.5 fm)$^3$ volume.
We have also discussed an equivalence of the method used in this study
to another derivative method~\cite{deDivitiis:2012vs}.

The results from the DFF method near the physical point 
are compared with the ones from 
the standard form factor calculation on the same volume ($L=64$) and also
on the larger volume ($L=128$) in our previous work~\cite{Shintani:2018ozy}.
For $\langle r_E^2\rangle$, $\mu$, $\langle r_M^2\rangle$, and 
$\langle r_A^2\rangle$,
the statistical uncertainties of the results in the DFF method are
relatively larger than those obtained from the fitting of the $q^2$ dependence of
the corresponding form factors. However, the DFF method can avoid the systematic error 
associated with the model dependence of the fit form.
Such a systematic error is dominant over the statistical error  
in the standard method, and then it makes the total accuracy worse than the DFF results.
We have also confirmed that the results from the DFF method are in good agreement with 
the ones from the standard analysis with the form factors within the combined errors
of the statistical and systematic uncertainties in both methods, except for the quantity of $F_P(0)$.

Two ways to determine $F_P(0)$ in the DFF method provide inconsistent results. 
A similar discrepancy is observed in our pilot calculation at a heavier pion mass of 
$M_\pi = 0.51$ GeV on the smaller volume of (2.9 fm)$^3$, while it becomes 
resolved on a larger volume of (5.8 fm)$^3$.
We thus have considered that the discrepancy comes from a finite volume effect,
and the significant finite volume effect can be related to 
a steep behavior of $F_P(q^2)$ near the origin, in other words, 
its broad shape in the coordinate space.

In our pilot calculation at a heavier pion mass of $M_\pi =0.51$ GeV,
we have also found that the DFF method
is more sensitive to finite volume effect than the standard method
as reported in the previous works~\cite{Lellouch:1994zu,Feng:2019geu}.
Therefore, an undetermined systematic error stemming from the finite volume effect
might exist in the DFF results from our numerical simulations on the (5.5 fm)$^3$ volume
near the physical point, although they agree with the results obtained by the standard method.
One of the important future works is a comprehensive study of systematic errors
in the DFF method, including the one from excited state contamination, 
at the physical point using the (10.9 fm)$^3$ volume
as is done in our previous study of the form factors 
in the standard method~\cite{Shintani:2018ozy}.
In this future direction, 
some improvements in the DFF method are needed to reduce the finite volume effect in the analysis level.
Such an improved analysis in the DFF method was proposed
for the case of meson form factors~\cite{Lellouch:1994zu,Feng:2019geu},
so that it can be extended and applied to the nucleon form factors.

\section*{Acknowledgments}
We thank members of the PACS collaboration for useful discussions.
Numerical calculations in this work were performed on Oakforest-PACS
in Joint Center for Advanced High Performance Computing (JCAHPC)
and Cygnus in Center for Computational Sciences at University of Tsukuba
under Multidisciplinary Cooperative Research Program of Center for Computational Sciences, University of Tsukuba, and Wisteria/BDEC-01 in the Information Technology Center, The University of Tokyo.
This research also used computational resources of the HPCI system provided by Information Technology Center of the University of Tokyo and RIKEN CCS through the HPCI System Research Project (Project ID: hp170022, hp180051, hp180072, hp180126, hp190025, hp190081, hp200062, hp210088, hp200188).
The calculation employed OpenQCD system\footnote{http://luscher.web.cern.ch/luscher/openQCD/}.
This work was supported in part by Grants-in-Aid 
for Scientific Research from the Ministry of Education, Culture, Sports, 
Science and Technology (Nos. 18K03605, 19H01892).

\appendix

\section{Momentum derivative under a partially quenched approximation}
\label{app:pq_approx}

In this appendix,
variables $x,y,z$ represent four-dimensional coordinates.
Let us consider the quark propagator $G_{p_j}$, which satisfies 
\begin{equation}
\sum_{z}D_{p_j}(x,z) G_{p_j}(z,y) = \delta_{x,y} .
\end{equation}
where $D_{p_j}$ represents the Dirac operator constructed with the gauge link
that is applied by the phase rotation associated with the momentum $p_j$ 
as $U_j(x) \to e^{ip_j} U_j(x)$~\cite{deDivitiis:2012vs}.
This phase rotation is nothing but a uniform external magnetic field. 
The expectation value of an observable ${\cal O}$
in the theory with a single quark field subjected to uniform phase rotation
is defined by
\begin{equation}
\langle {\mathcal O} \rangle_{p_j} =
\frac{\int {\mathcal D}U {\mathcal O} \det D_{p_j}\, e^{-S_{\rm eff}(U)}}{Z(p_j)}
\label{eq:exp_Dp}
\end{equation}
with
\begin{equation}
Z(p_j) = \int {\mathcal D}U \det D_{p_j}\, e^{-S_{\rm eff}(U)} .
\label{eq:Z_Dp}
\end{equation}
where
$S_{\rm eff}(U)$ contains the gauge action and also may contain the term
associated with the determinants of the Dirac operator for other quarks, which are independent of $p_j$.
Using the above definition, a derivative of $\langle G_{p_j}(x,y) \rangle_{p_j}$ 
with respect to $p_j$ at zero momentum is expressed 
by three terms\footnote{Although the third term should be vanished because of $\langle {\rm Tr}\left[G(z,z) \widetilde{\gamma}_j(z)\right]\rangle= 0$, we write down it explicitly. } as,
\begin{eqnarray}
\left.\frac{\partial}{\partial p_j}\langle G_{p_j}(x,y) \rangle_{p_j}\right|_{{\bf p}=0} &=& 
-i\sum_z\left\langle G(x,z)\widetilde{\gamma}_j(z)G(z,y)\right\rangle\nonumber\\
&&
+i\sum_{z}\left\langle {\rm Tr}\left[G(z,z) \widetilde{\gamma}_j(z) \right] 
G(x,y) \right\rangle
-i\sum_{z}\left\langle {\rm Tr}\left[G(z,z) \widetilde{\gamma}_j(z)\right]\right\rangle
\langle G(x,y)\rangle ,\nonumber\\
\label{eq:der_mom_full_obs}
\end{eqnarray}
where we use the following relation
\begin{equation}
\left.\frac{\partial}{\partial p_j}\det D_{p_j}\right|_{{\bf p}=0} = 
i\det D \sum_z {\rm Tr}
\left[G(z,z) \widetilde{\gamma}_j(z)\right] ,
\end{equation}
with
\begin{equation}
\left.\frac{\partial}{\partial p_j} D_{p_j}(x,y)
\right|_{{\bf p}=0} 
= i \widetilde{\gamma}_j (x) \delta_{x,y}.
\end{equation}
The right hand side of Eq.~(\ref{eq:der_mom_full_obs}) is nothing but 
the left hand side of Eq.~(\ref{eq:equiv_mom_der}) after the Wick contraction.

Under a partially quenched approximation of the quark field subjected to uniform phase rotation,
the condition of $\det D_{p_j} = 1$ is imposed in Eqs.~(\ref{eq:exp_Dp}) and (\ref{eq:Z_Dp}). 
It thus ends up that the second and third terms in Eq.~(\ref{eq:der_mom_full_obs}) disappear, 
since both terms arise from the derivative of $\det D_{p_j}$.
Therefore, the momentum derivative of the quark propagator is
expressed only by the first term in Eq.~(\ref{eq:der_mom_full_obs})
{\it under the partially quenched approximation}.

\section{Result of form factors on a $64^4$ lattice}
\label{app:form_factors}

In this appendix, 
the results for the form factors calculated near 
the physical point on the $L=64$ lattice volume
are summarized.
The momentum transfer squared $q^2$ is calculated by
$q^2 = 2 M_N(E_N(p)-M_N)$ with $E_N(p) = \sqrt{M_N^2 + p^2}$.
The energy $E_N(p)$ is determined using the measured $M_N$ 
and lattice momentum ${\bf p} = 2\pi/64 \times {\bf n}$ 
with integer vectors ${\bf n}$.
The values of $q^2$ in each momentum used in this study
are listed in Table~\ref{tab:mom}.
The form factors are evaluated from the ratios of 
Eqs.~(\ref{eq:R_GE})--(\ref{eq:R_FA_FP}) in the asymptotic region,
whose values are determined from a constant fit with the fitting range of
$t = 4$--8 for $t_{\rm sep} = 12$, $t = 5$--9 for $t_{\rm sep}=14$,
and $t = 5$--11 for $t_{\rm sep}=16$.

\begin{figure}[!ht]
 \centering
 \includegraphics*[scale=0.35]{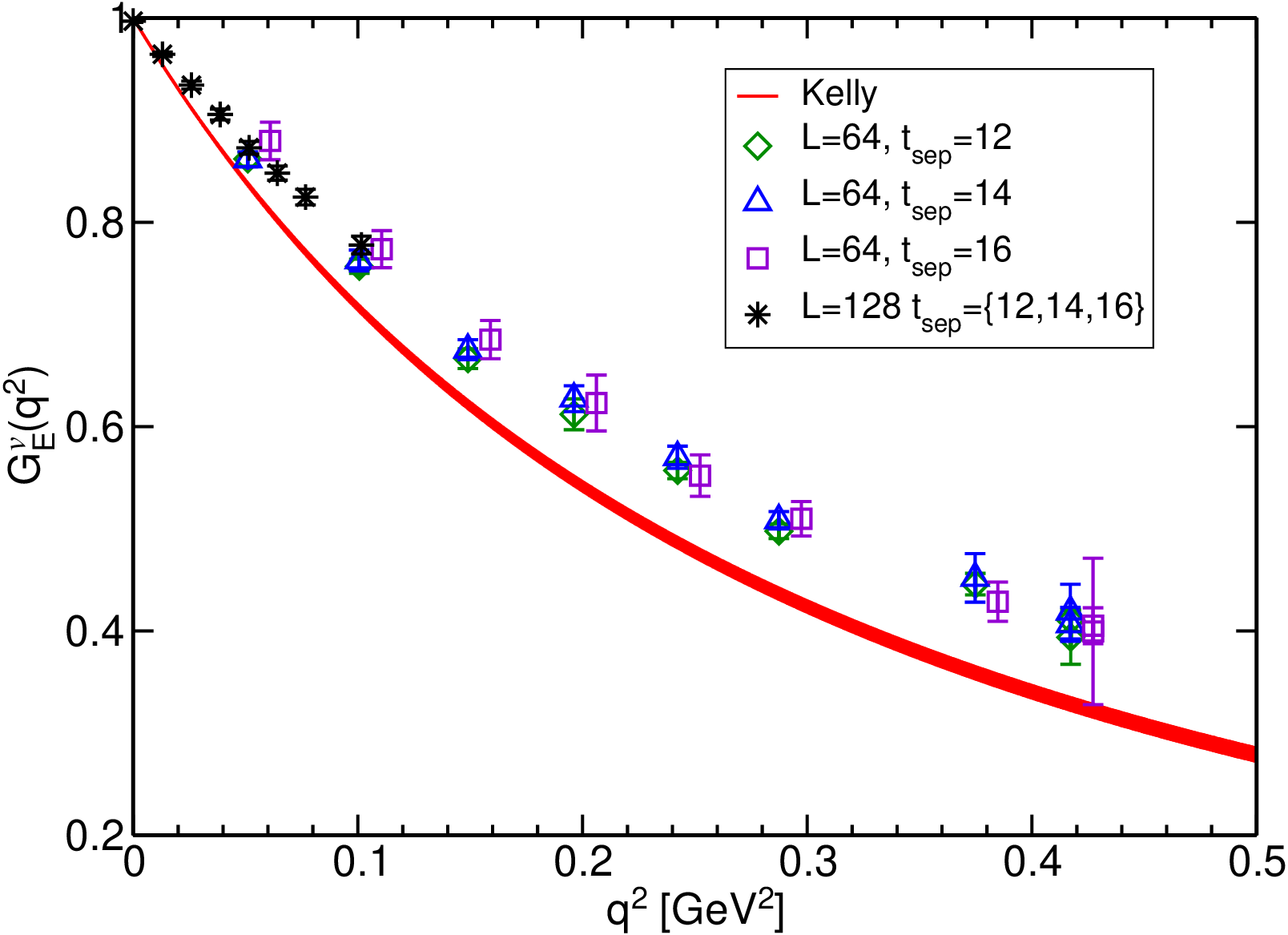}
 \caption{Result of $G_E(q^2)$ with $t_{\rm sep}=12$, 14, and 16 on the $L=64$ lattice volume as a function of $q^2$.
The data of $t_{\rm sep}=16$ are slightly shifted to the positive 
$x$ direction for clarity.
Our previous result on the $L=128$ volume~\cite{Shintani:2018ozy} 
given after taking average of three data sets calculated with $t_{\rm sep}=12$, 14 and 16
is also plotted by the asterisk symbol.
The red curve represents Kelly's parametrization of the experiment data~\cite{Kelly:2004hm}.
  \label{fig:GE_64}
 }
\end{figure}

\begin{figure}[!ht]
 \centering
 \includegraphics*[scale=0.35]{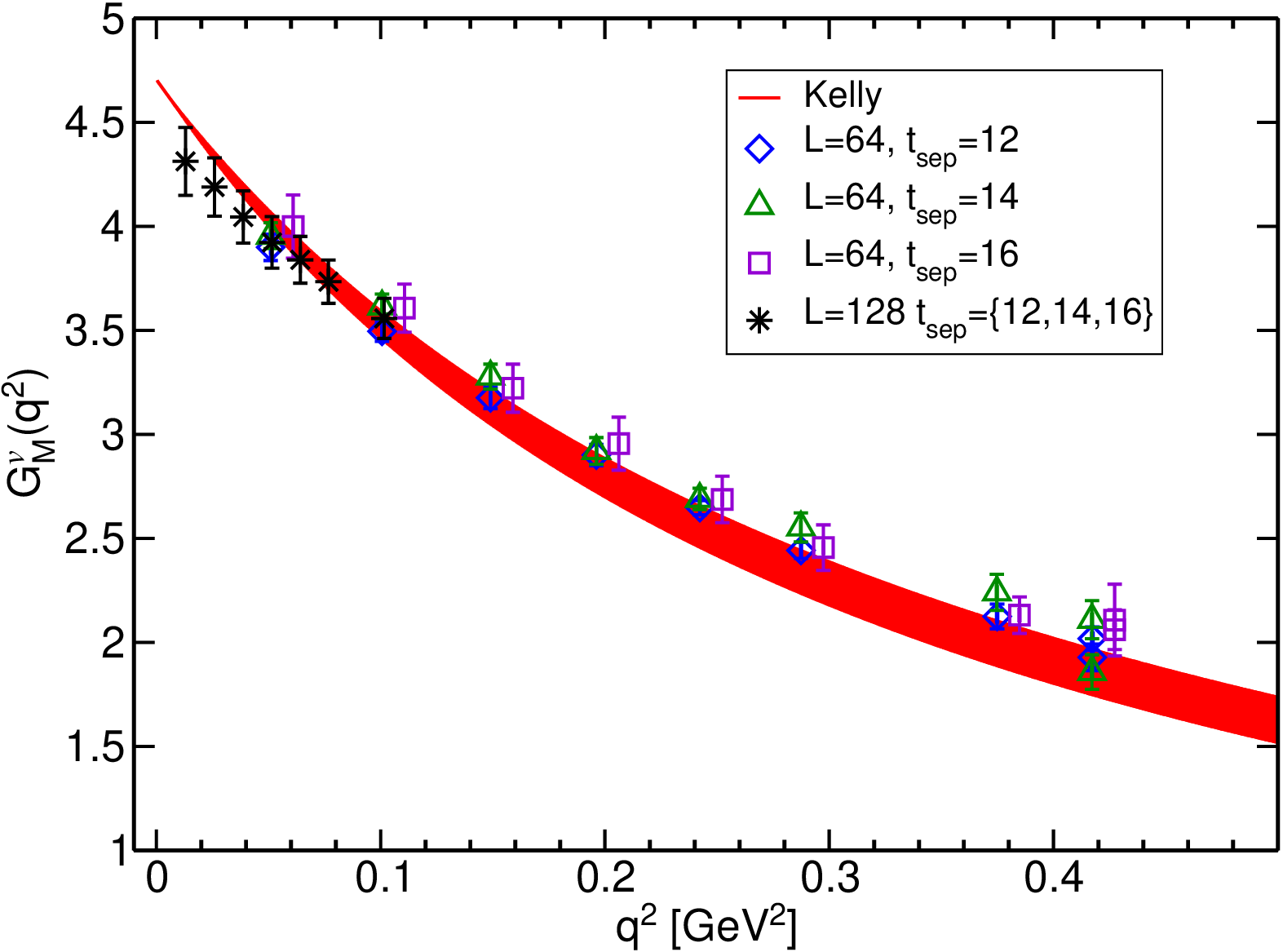}
 \caption{Same as Fig.~\ref{fig:GE_64} for $G_M(q^2)$.
 The red curve represents Kelly's parametrization of the experiment data~\cite{Kelly:2004hm}.
  \label{fig:GM_64}
 }
\end{figure}

\begin{figure}[!ht]
 \centering
 \includegraphics*[scale=0.35]{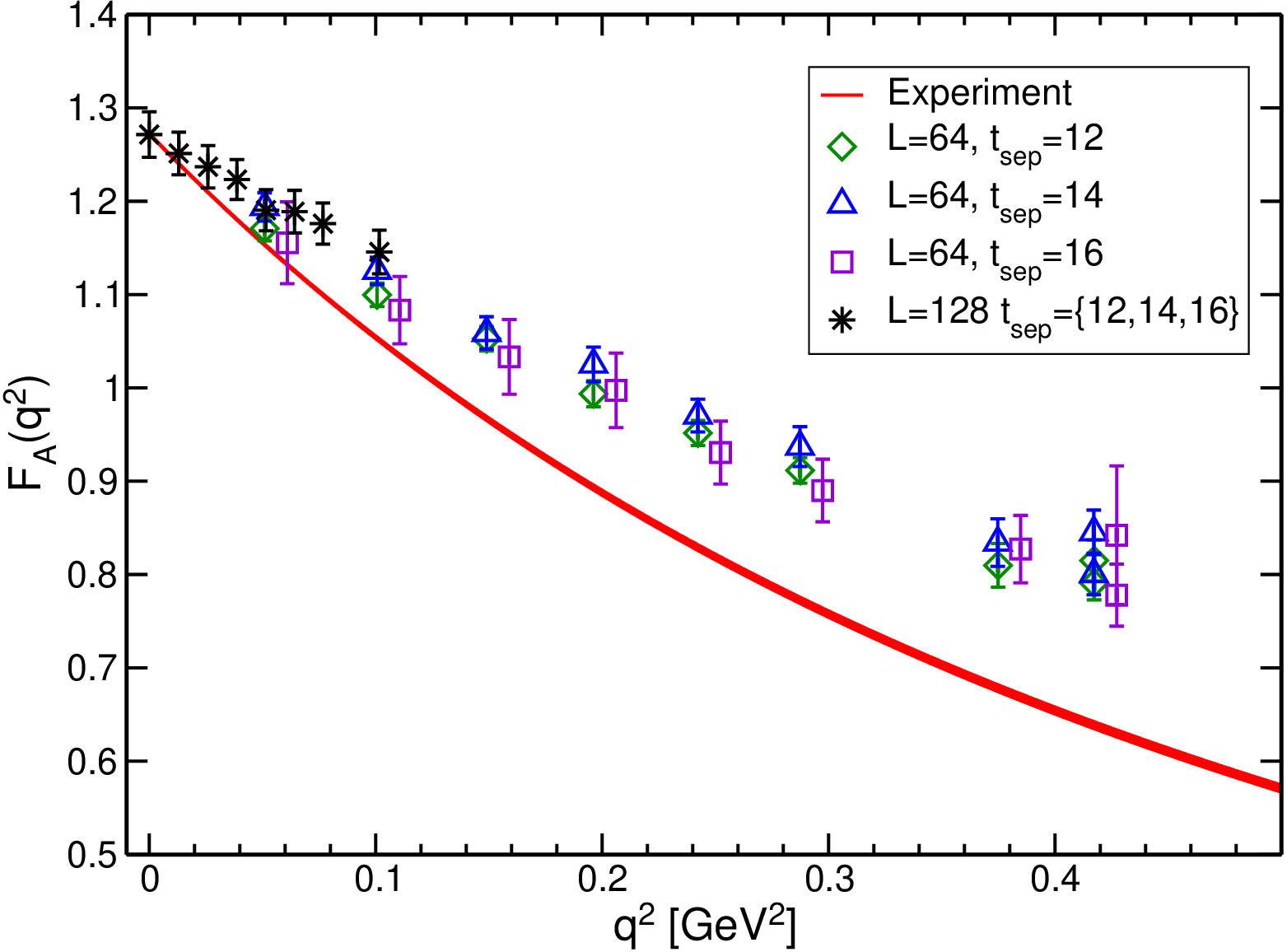}
 \caption{Same as Fig.~\ref{fig:GE_64} for $F_A(q^2)$.
The red curve is given by a dipole form with the dipole mass~\cite{Bodek:2007ym,Bernard:2001rs} 
and $g_A$~\cite{Zyla:2020zbs}.
  \label{fig:FA_64}
 }
\end{figure}

\begin{figure}[!ht]
 \centering
 \includegraphics*[scale=0.35]{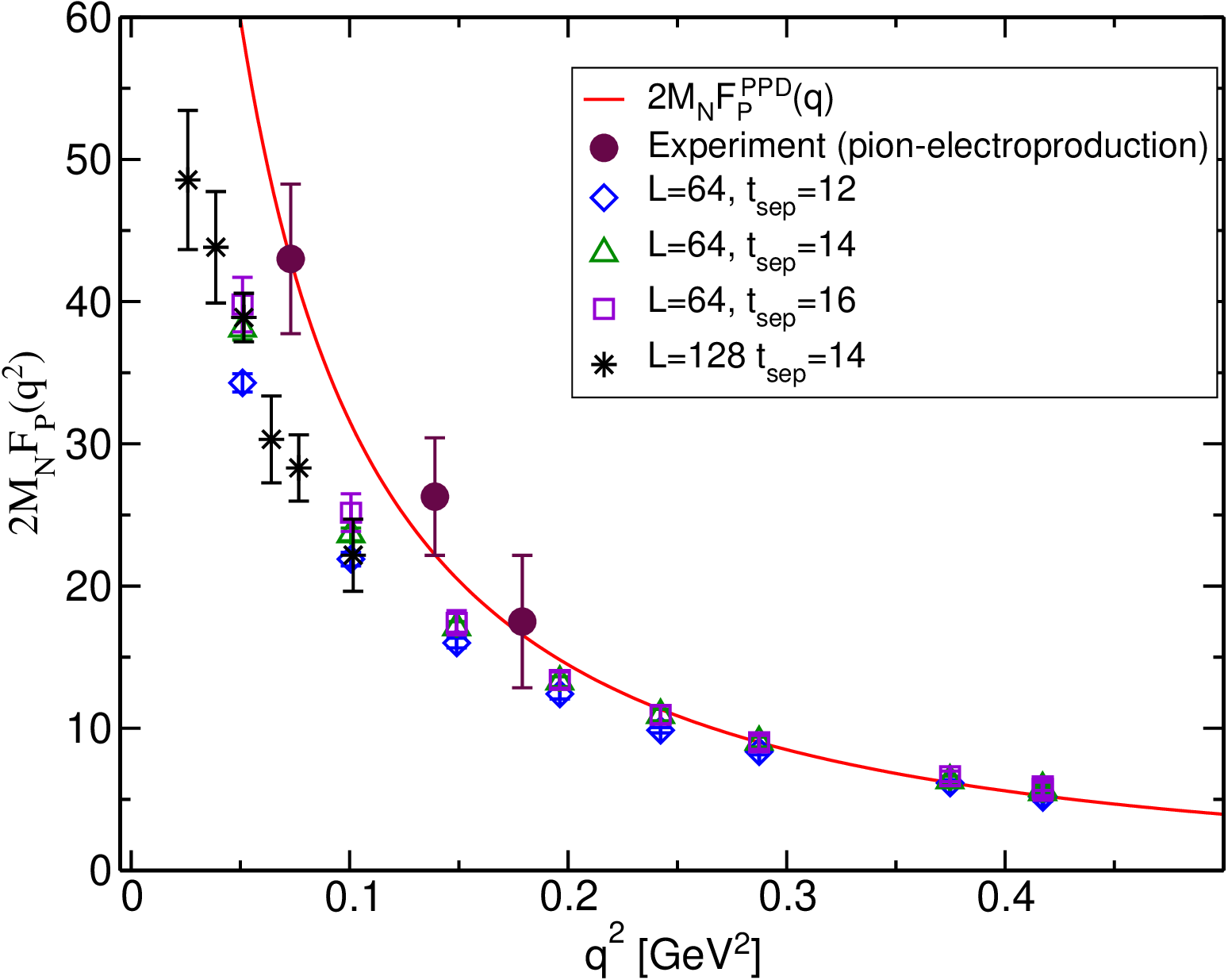}
 \caption{Same as Fig.~\ref{fig:GE_64} for $2M_NF_P(q^2)$.
The red curve is given by the pion-pole dominance model (PPD) 
$2M_N F_P^{\rm PPD}(q^2) = 4M_N^2 F_A(q^2)/(M_\pi^2 + q^2)$ with a dipole 
form of $F_A(q^2)$ using the dipole mass~\cite{Bodek:2007ym,Bernard:2001rs} 
and $g_A$~\cite{Zyla:2020zbs}.
The experimental result of the pion-electroproduction~\cite{Choi:1993vt}
is also plotted by the diamond symbol.
 \label{fig:FP_64}
 }
\end{figure}

The results for the renormalized isovector 
$G_E(q^2)$, $G_M(q^2)$, $F_A(q^2)$, and $F_P(q^2)$ with 
$t_{\rm sep}=12$, 14, and 16 are shown in 
Figs.~\ref{fig:GE_64}--\ref{fig:FP_64} together with
those obtained in the previous calculation on
a 128$^4$ lattice~\cite{Shintani:2018ozy} for comparison.
For $F_A(q^2)$ and $F_P(q^2)$, the error of 
$Z_A=0.9650(68)$~\cite{Ishikawa:2015fzw} is included in their errors.
The values of each form factor obtained with $t_{\rm sep}=12$, 14, and 16
are tabulated in Table~\ref{tab:ff_tsep12}, \ref{tab:ff_tsep14}, and \ref{tab:ff_tsep16}, respectively.
Our results with $t_{\rm sep}=12$, 14, and 16 on the $L=64$ lattice volume are consistent
with each other, and also statistically agree 
with the data on the $L=128$ lattice volume~\cite{Shintani:2018ozy}
in all the form factors, except for $F_P(q^2)$.
There is a clear discrepancy between the data of $F_P(q^2)$ 
obtained with $t_{\rm sep} = 12$ and 14 on the $L=64$ lattice volume.
It is considered to be caused by significant effect from excited state 
contamination as reported in Refs.~\cite{Shintani:2018ozy,Bar:2018xyi,Jang:2019vkm,Alexandrou:2020okk,Jang:2019jkn,Park:2021ypf}.

\begin{figure}[!ht]
 \centering
 \includegraphics*[scale=0.4]{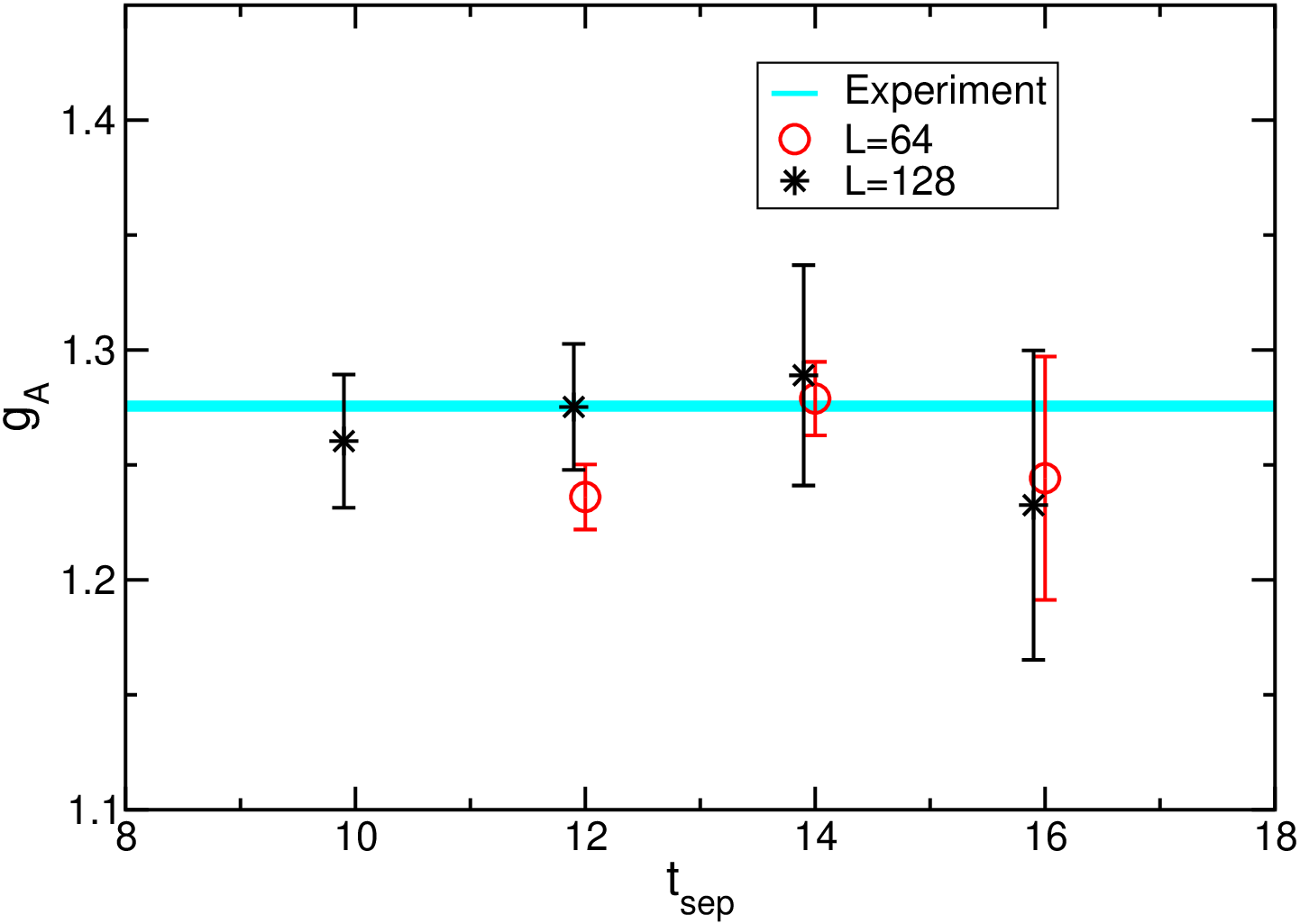}
\caption{The renormalized $g_A$ from the $L=64$ and $L=128$~\cite{Shintani:2018ozy} lattices as a function of $t_{\rm sep}$.
The data of $L=128$ are slightly shifted to the negative
$x$ direction for clarity.
The cyan band represents the experimental value~\cite{Zyla:2020zbs}.
  \label{fig:gA_64_128}
 }
\end{figure}

It is worth remarking that the value of $g_A$ obtained with $t_{\rm sep}=12$ 
is slightly smaller than the one obtained with $t_{\rm sep}=14$.
Since the data with $t_{\rm sep} = 16$ has a much larger statistical error
than the two data,
the $t_{\rm sep}$ dependence is unclear as shown in Fig.~\ref{fig:gA_64_128}.
The figure also shows that any appreciable $t_{\rm sep}$ dependence 
was not observed in the range of $t_{\rm sep}=10$--16 in the larger volume 
calculation~\cite{Shintani:2018ozy}, and they are statistically consistent
with all the data on the $L=64$ lattice.
Therefore, it is not clear whether this slight difference between two results from $t_{\rm sep}=12$ and 14
is just a statistical fluctuation or related to the systematic errors, {\it e.g.,} the finite volume effect and the excited state contamination.
To clarify this point, a further systematic study using a large set of different $t_{\rm sep}$ with the statistical error as small as 
the two data is needed.

The fit results with dipole, quadratic, and z-expansion functions
in Eqs.~(\ref{eq:dipole_form})--(\ref{eq:z-exp_form})
for $G_E(q^2)$, $G_M(q^2)$, and $F_A(q^2)/g_A$ are
summarized in Tables~\ref{tab:fit_GE}--\ref{tab:fit_FA}.
The value of $\chi^2/$dof is obtained from a correlated fit.
The maximum $q^2$ in each fit denoted by $q_{\rm cut}^2$ in the tables is chosen
to obtain an acceptable $\chi^2/$dof.
In all the fits for $G_E(q^2)$ and $F_A(q^2)/g_A$, 
we use the condition of $G_E(0) = F_A(0)/g_A = 1$ imposed in the respective fit functions.

The central values and their statistical errors presented in Table~\ref{tab:results_der}
are determined by the dipole fits of the respective form factors.
Their systematic errors are estimated from the maximum discrepancy
from the dipole fit result with the two other fits.
Note that a cubic z-expansion fit of $G_M(q^2)$ 
gives a negative value of the MS magnetic radius in contrast to the results from 
other two fits. In our data of $G_M(q^2)$, the z-expansion fit indeed becomes unstable
once higher powers of $z$ are included, while a quadratic z-expansion fit for $G_M(q^2)$ gives a positive radius.
Nevertheless, in estimate of the systematic error of $G_M(q^2)$,
we use the result obtain from the cubic z-expansion fit of $G_M(q^2)$ to hold
the same evaluation used in the other form factors.

\begin{table}[!ht]
\caption{List of integer vectors ${\bf n}_i$ for 
the momentum ${\bf p}_i = 2\pi{\bf n}_i/L$ (with $L=64$)
projected on the nucleon 2-point and 3-point functions.
The degeneracy in each ${\bf n}_i$ ($N_{\rm deg}$),
and the corresponding values of the momentum transfer squared $q^2_i = 2M_N(E_N(p_i)-M_N)$ 
are also tabulated.
\label{tab:mom}
}
\begin{ruledtabular}
\begin{tabular}{ccccccccccc} 
$i$ & 0 & 1 & 2 & 3 & 4 & 5 & 6 & 7 & 8 & 9 \\\hline
${\bf n}_i$ & $(0,0,0)$ & $(1,0,0)$ & $(1,1,0)$ & $(1,1,1)$ & $(2,0,0)$ &
$(2,1,0)$ & $(2,1,1)$ & $(2,2,0)$ & $(2,2,1)$ & $(3,0,0)$\\ 
$N_{\rm deg}$ & 1 & 6 & 12 & 8 & 6 & 24 & 24 & 12 & 24 & 6\\
$q_i^2$ [GeV$^2$] & 0 & 0.051 & 0.101 & 0.149 & 0.196 & 0.242 & 0.288 &
0.375 & 0.418 & 0.418 \\
\end{tabular}
\end{ruledtabular} 
\end{table}

\begin{table*}[ht!]
\caption{Results of the form factors with $t_{\rm sep}=12$.
\label{tab:ff_tsep12}
}
\begin{ruledtabular}
\begin{tabular}{cccccc} 
$i$ & $q_i^2\,[{\rm GeV}^2]$ & $G_E(q^2)$ & $G_M(q^2)$ & $F_A(q^2)$ & $2M_NF_P(q^2)$ \\  
  \hline
0& 0.000 & 1.0000     & ---       & 1.236(14) & --- \\
1& 0.051 & 0.8620(68) & 3.899(63) & 1.171(13) & 34.29(64)\\
2& 0.101 & 0.7569(70) & 3.495(48) & 1.100(13) & 21.90(48)\\
3& 0.149 & 0.6664(96) & 3.175(53) & 1.053(13) & 16.00(34)\\
4& 0.196 & 0.612(15)  & 2.900(52) & 0.994(14) & 12.42(36)\\
5& 0.242 & 0.5570(80) & 2.647(34) & 0.952(13) & 9.86(19)\\
6& 0.288 & 0.4976(73) & 2.442(38) & 0.912(14) & 8.35(22)\\
7& 0.375 & 0.446(11)  & 2.125(60) & 0.810(23) & 6.15(21) \\
8& 0.418 & 0.410(11)  & 2.018(56) & 0.791(18) & 5.22(21) \\
9& 0.418 & 0.394(26)  & 1.928(64) & 0.815(22) & 5.14(26) \\
\end{tabular}
\end{ruledtabular} 
\end{table*}

\begin{table*}[ht!]
\caption{Results of the form factors with $t_{\rm sep}=14$.
\label{tab:ff_tsep14}
}
\begin{ruledtabular}
\begin{tabular}{cccccc} 
$i$ & $q_i^2\,[{\rm GeV}^2]$ & $G_E(q^2)$ & $G_M(q^2)$ & $F_A(q^2)$ & $2M_N F_P(q^2)$ \\  
  \hline
0& 0.000 & 1.0000    & ---       & 1.279(16) & --- \\
1& 0.051 & 0.8616(74)& 3.955(62) & 1.194(15) & 38.09(78)\\
2& 0.101 & 0.763(10) & 3.616(59) & 1.125(15) & 23.63(45)\\
3& 0.149 & 0.6752(98)& 3.277(60) & 1.059(17) & 17.09(42)\\
4& 0.196 & 0.627(13) & 2.920(52) & 1.025(19) & 13.27(35)\\
5& 0.242 & 0.570(11) & 2.689(52) & 0.970(18) & 10.93(40)\\
6& 0.288 & 0.5082(86)& 2.552(70) & 0.937(21) & 9.03(34)\\
7& 0.375 & 0.452(24) & 2.240(87) & 0.834(25) & 6.33(31) \\
8& 0.418 & 0.406(17) & 2.109(92) & 0.801(22) & 5.79(27) \\
9& 0.418 & 0.419(27) & 1.857(83) & 0.845(24) & 5.49(25) \\
\end{tabular}
\end{ruledtabular} 
\end{table*}

\begin{table*}[ht!]
\caption{Results of the form factors with $t_{\rm sep}=16$.
\label{tab:ff_tsep16}
}
\begin{ruledtabular}
\begin{tabular}{cccccc} 
$i$ & $q_i^2\,[{\rm GeV}^2]$ & $G_E(q^2)$ & $G_M(q^2)$ & $F_A(q^2)$ & $2M_N F_P(q^2)$ \\  
  \hline
0& 0.000 & 1.0000    & ---      & 1.244(53) & --- \\
1& 0.051 & 0.880(18) & 4.00(15) & 1.155(44) & 39.8(1.9)\\
2& 0.101 & 0.774(18) & 3.61(12) & 1.083(36) & 25.2(1.3)\\
3& 0.149 & 0.685(19) & 3.22(16) & 1.033(40) & 17.41(85)\\
4& 0.196 & 0.623(27) & 2.96(13) & 0.997(40) & 13.40(57)\\
5& 0.242 & 0.552(20) & 2.69(11) & 0.931(34) & 10.94(63)\\
6& 0.288 & 0.510(17) & 2.46(11) & 0.890(33) & 9.01(49)\\
7& 0.375 & 0.429(19) & 2.131(88)& 0.827(36) & 6.62(37) \\
8& 0.418 & 0.405(18) & 2.061(95)& 0.778(33) & 5.92(29) \\
9& 0.418 & 0.399(72) & 2.11(17) & 0.842(74) & 5.74(80) \\
\end{tabular}
\end{ruledtabular} 
\end{table*}

\begin{table}[!ht]
\caption{Fit results of $G_E(q^2)$ with $t_{\rm sep}=14$ using 
dipole, quadratic, and z-expansion
forms in Eqs.~(\ref{eq:dipole_form})--(\ref{eq:z-exp_form}).
The fit range is $q^2 = q_1^2$--$q_{\rm cut}^2$.
In all the fit we employ $G_E(0) = 1$.
\label{tab:fit_GE}
}
\begin{ruledtabular}
\begin{tabular}{cccc}
 & dipole & quadratic & z-exp(cubic) \\\hline
$\langle r_E^2 \rangle$ [fm$^2$] & 0.646(16) & 0.624(26) & 0.737(62) \\
$\chi^2$/dof & 1.7(0.9) & 2.8(1.9) & 1.3(0.9) \\
$q^2_{\rm cut}$ [GeV$^2$] & 0.418 & 0.242 & 0.418 \\
\end{tabular}
\end{ruledtabular} 
\end{table}

\begin{table}[!ht]
\caption{Fit results of $G_M(q^2)$ with $t_{\rm sep}=14$ using
dipole, quadratic, and z-expansion
forms in Eqs.~(\ref{eq:dipole_form})--(\ref{eq:z-exp_form}).
The fit range is $q^2 = q_1^2$--$q_{\rm cut}^2$.
The fit result with a quadratic z-expansion form is also tabulated.
\label{tab:fit_GM}
}
\begin{ruledtabular}
\begin{tabular}{ccccc}
 & dipole & quadratic & z-exp(cubic) & z-exp(quadratic)\\\hline
$\mu$ & 4.357(42) & 4.316(58) & 3.53(28) & 4.31(11) \\
$\langle r_M^2 \rangle$ [fm$^2$] & 0.579(20) & 0.489(17) & $-$2.1(1.1) & 0.211(16) \\
$\chi^2$/dof & 2.7(1.2) & 2.1(1.2) & 2.4(2.3) & 1.0(1.4) \\
$q^2_{\rm cut}$ [GeV$^2$] & 0.418 & 0.418 & 0.288 & 0.242 \\
\end{tabular}
\end{ruledtabular} 
\end{table}

\begin{table}[!ht]
\caption{Fit results of $F_A(q^2)/g_A$ with $t_{\rm sep}=14$ using 
dipole, quadratic, and z-expansion
forms in Eqs.~(\ref{eq:dipole_form})--(\ref{eq:z-exp_form}).
The fit range is $q^2 = q_1^2$--$q_{\rm cut}^2$.
In all the fit we employ $F_A(0)/g_A = 1$.
\label{tab:fit_FA}
}
\begin{ruledtabular}
\begin{tabular}{cccc}
 & dipole & quadratic & z-exp(cubic) \\\hline
$\langle r_A^2 \rangle$ [fm$^2$] & 0.308(17) & 0.323(19) & 0.337(39) \\
$\chi^2$/dof & 0.4(0.9) & 1.4(1.2) & 1.6(1.5) \\
$q^2_{\rm cut}$ [GeV$^2$] & 0.149 & 0.288 & 0.288 \\
\end{tabular}
\end{ruledtabular} 
\end{table}

\clearpage
\bibliographystyle{apsrev4-1}
\bibliography{ref}
\end{document}